\documentclass[twocolumn,showpacs,preprintnumbers,amsmath,amssymb,nofootinbib ]{revtex4}

\usepackage{graphicx}
\usepackage{dcolumn}
\usepackage{bm}
\usepackage{epsfig}
\usepackage{psfrag}

\usepackage{dsfont}
\usepackage{cancel}
\usepackage{graphicx}

\begin{document}

\title{Electroweak form factors of heavy-light mesons
--\\a relativistic point-form approach}
\author{Mar\'ia G\'omez-Rocha}
\email{maria.gomez-rocha@uni-graz.at}
\author{Wolfgang Schweiger}%
\email{wolfgang.schweiger@uni-graz.at} \affiliation{ Institut
f\"ur Physik, Universit\"at Graz, A-8010 Graz, Austria}
\date{\today}

\date{\today}

\begin{abstract}
We present a general relativistic framework for the calculation of
the electroweak structure of heavy-light mesons within
constituent-quark models. To this aim the physical processes in
which the structure is measured, i.e. electron-meson scattering and
semileptonic weak decays, are treated in a Poincar\'e invariant way
by making use of the point-form of relativistic quantum mechanics.
The electromagnetic and weak meson currents are extracted from the
$1$-$\gamma$ and $1$-$W$-exchange amplitudes that result from a
Bakamjian-Thomas type mass operator for the respective systems. The
covariant decomposition of these currents provides the
electromagnetic and weak (transition) form factors. Problems with
cluster separability, which are inherent in the Bakamjian-Thomas
construction, are discussed and it is shown how to keep them under
control. It is proved that the heavy-quark limit of the electroweak
form factors leads to one universal function, the Isgur-Wise
function, confirming that the requirements of heavy-quark symmetry
are satisfied. A simple analytical expression is given for the
Isgur-Wise function and its agreement with a corresponding
front-form calculation is verified numerically. Electromagnetic form
factors for $B^-$ and $D^+$ and weak $B\rightarrow D^{(\ast)}$-decay
form factors are calculated with a simple harmonic-oscilllator wave
function and heavy-quark symmetry breaking due to finite masses of
the heavy quarks is discussed.
\end{abstract}

\pacs{13.40.Gp, 11.80.Gw, 12.39.Ki, 14.40.Aq}

\maketitle

\section{\label{sec:introduction} Introduction}
A proper relativistic formulation of the electroweak structure of
few-body bound states poses several problems. Even if one has model
wave functions for the few-body bound states one is interested in,
it is not straightforward to construct electromagnetic and weak
currents with all the properties it should have. Two basic features
are Poincar\'e covariance and cluster
separability~\cite{Sokolov:1977ym,Coester:1982vt,Keister:1991sb}.
The latter means that the bound-state current should become a sum of
subsystem currents, if the interaction between the subsystems is
turned off. This property is closely related to the requirement that
the charge of the whole system should be the sum of the subsystem
charges, irrespective whether the interaction is present or
not~\cite{lev:1995}. Electromagnetic currents should, furthermore,
satisfy current conservation and in the case of electroweak currents
of heavy-light systems one has restrictions coming from heavy-quark
symmetry that should be satisfied if the mass of the heavy quark
goes to infinity~\cite{Isgur:1989vq,Isgur:1989ed,Neubert:1993mb}.
This is the topic which we will concentrate on in this paper
keeping, of course, also the other requirements for a reasonable
current in mind.

The main ingredients in the construction of currents are the wave
functions of the incoming and outgoing few body bound states. Since
momentum is transferred to the bound state in the course of an
electroweak process, one has to know how to boost the wave function,
which is usually calculated for the bound state at rest, to the
initial and final state, respectively. A procedure which provides
wave functions for interacting few-body systems with well defined
relativistic boost properties is the, so called, Bakamjian-Thomas
construction~\cite{Bakamjian:1953kh,Keister:1991sb}. It gives an
interacting representation of the Poincar\'e algebra on a few-body
Hilbert space, allows even for instantaneous interactions, and it
works in the 3 common forms of relativistic Hamiltonian
dynamics~\cite{Dirac:1949cp}, the instant form, the front form and
the point form. These forms are characterized by which of the
Poincar\'e generators contain interaction terms  and which are
interaction free. In the point-form, which we are going to use, all
4 components of the 4-momentum operator are interaction dependent,
whereas the Lorentz generators stay free of interactions. As a
consequence boosts and the addition of angular momenta become
simple.

There is a long list of papers in which relativistic constituent-quark models serve as a starting point for the calculation of the electroweak structure of heavy-light mesons. A lot of these calculations have been done in front form, like e.g. those in Refs.\cite{Jaus:1996np,Simula:1996pk,Demchuk:1995zx,Cheng:1996if} , to mention a few. In these papers the electromagnetic and weak meson currents are usually approximated by one-body currents, which means that those currents are assumed to be a sum of contributions in which the gauge boson couples only to one of the constituents, whereas the others act as spectators. It is well known that this approximation leads to problems with covariance of the currents in front form and in instant form~\cite{lev:1995}. The form factors extracted from such a one-body approximation of a current depend, in general, on the frame in which the approximation is made. In the covariant front-form formulation suggested in Ref.~\cite{Carbonell:1998rj} this problem is circumvented by introducing additional, spurious covariants and form factors that are associated with the chosen orientation of the light front. Another way to cure this problem is the introduction of a non-valence contribution leading to a, so called, Z-graph~\cite{Simula:2002vm,Bakker:2003up}. Such a non-valence contribution to the currents is also included in an effective way in the instant-form approach of Ebert et al.~\cite{Ebert:2006nz}. This is a very sophisticated constituent-quark model for heavy-light systems based on a quasipotential approach. A whole series of papers by Ebert and collaborators deals very comprehensively with spectroscopy, structure and decays of heavy-light mesons and baryons. In connection with instant form constituent-quark models one should also mention the papers of Le Yaouanc et al. (see, e.g., Ref.~\cite{Morenas:1997nk} and references therein). They were the first to prove that covariance of a one-body current is recovered, if the mass of the heavy quark goes to infinity~\cite{LeYaouanc:1995wv}. Thereby they made use of the known boost properties of wave functions within the Bakamjian-Thomas formulation of relativistic quantum mechanics.

On the contrary, the literature on point-form calculations of heavy-light systems is very sparse, although the point form seems to be particularly suited for the treatment of this kind of systems. We are only aware of two papers by Keister~\cite{Keister:1992wq,Keister:1997je}. This is one of our motivations to investigate the electroweak structure of heavy-light mesons within the point form of relativistic dynamics. Although it is possible to formulate a covariant one-body current in point form~\cite{Klink:1998qf,Melde:2004qu}, we will adopt a different strategy. Instead of making a particular ansatz for the electromagnetic and weak currents and extract the form factors from these currents, we rather want to derive these currents in such a way that they are compatible with the binding forces. The idea is to treat the physical processes in which the electroweak form factors are measured in a Poincar\'e invariant way by means of the Bakamjian-Thomas formalism. This gives us 1-$\gamma$ and 1-$W$-exchange amplitudes from which the currents and form factors can be extracted. This kind of procedure has already been applied successfully to the calculate electromagnetic form factors of spin-0 and spin-1 two-body bound states consisting of equal-mass particles~\cite{Biernat:2009my,Biernat:2011mp}. These calculations were restricted to space-like momentum transfers. For instantaneous binding forces the results were found to be equivalent with those obtained with a one-body ansatz for the current in the covariant front-form approach~\cite{Carbonell:1998rj}. The present paper is an extension of the foregoing work to unequal-mass constituents and to weak decay form factors in the time-like momentum transfer region. It is also intended as a check whether the additional restrictions coming from heavy-quark symmetry can be accounted for within our approach.

The general Poincar\'e invariant framework that we use to describe
electron-meson scattering and semileptonic weak decays of mesons
will be introduced in Sec.~\ref{sec:formalism}. It is a relativistic
multichannel formalism for a Bakamjian-Thomas type mass
operator~\cite{Bakamjian:1953kh,Keister:1991sb} that is represented
in a velocity-state basis~\cite{Klink:1998qf}. This multichannel
formulation is necessary to account for the dynamics of $\gamma$-
and $W$-exchange, respectively. The one-photon-exchange amplitude
for electron scattering off a confined quark-antiquark pair is
derived in Sec.~\ref{subsec:scattering}, the one-$W$-exchange
amplitude for the semileptonic decay of a confined quark-antiquark
state into another (confined) quark-antiquark state in
Sec.~\ref{subsec:decay}. Since these amplitudes have the usual
structure, namely lepton current contracted with hadron current
times gauge-boson  propagator, it is easy to identify the
electromagnetic and weak hadron currents. This is explicitly done
for pseudoscalar mesons and pseodscalar-to-pseudoscalar as well as
pseudoscalar-to-vector transitions assuming that the mesons are pure
$s$-wave. The Lorentz-structure of the resulting electromagnetic and
weak currents is then analyzed in Sec.~\ref{sec:currents}. As a
result of this analysis we obtain the electroweak form factors.
Section~\ref{subsec:emff} contains also a short discussion of
cluster problems, connected with the Bakamjian-Thomas construction,
and their effect on the electromagnetic current. The limit of
heavy-quark mass going to infinity is investigated in
Sec.~\ref{sec:heavyqlim}. The precise definition of the \lq\lq
heavy-quark limit\rq\rq\ is introduced and it is proved that the
heavy-quark limit of the electromagnetic and weak from factors
yields a single universal function, the Isgur-Wise function. Model
calculations of the electromagnetic $D^+$ and $B^-$ form factors and
weak $B\rightarrow D^{(\ast)}$ decay form factors for physical
masses of the heavy quarks are presented in Sec.~\ref{sec:results}.
These are contrasted with the Isgur-Wise function to estimate
heavy-quark-symmetry breaking effects due to finite masses of the
heavy quarks. Our summary and conclusions are finally given in
Sec.~\ref{sec:conclusion}.

\section{\label{sec:formalism}Coupled-channel formalism}

\subsection{\label{subsec:formalism}Prerequisites}
In the point-form version of the Bakamjian-Thomas construction the
4-momentum operator for an interacting few-body system is written
as a product of an interaction-dependent mass operator times a
free 4-velocity operator~\cite{Keister:1991sb},
\begin{equation}
\hat{P}^\mu =
\hat{M}\,\hat{V}^\mu_\mathrm{free}=(\hat{M}_{\mathrm{free}}
+\hat{M}_\mathrm{int}) \, \hat{V}^\mu_{\mathrm{free}}\, .
\label{eq:momentumop}\end{equation}
Relativistic invariance holds if the interaction term
$\hat{M}_\mathrm{int}$ is a Lorentz scalar and commutes with
$\hat{V}^\mu_{\mathrm{free}}$. Equation~(\ref{eq:momentumop})
implies that the overall velocity of the system can be easily
separated from the internal motion and one can concentrate on
studying the mass operator $\hat{M}$ which is a function of the
internal variables only.

In this type of approach the operators of interest are most
conveniently represented in a velocity-state
basis~\cite{Klink:1998zz}. An $n$-particle velocity state $\vert
v; \vec{k}_1, \mu_1; \vec{k}_2, \mu_2;\dots; \vec{k}_n, \mu_n
\rangle $ is just a multiparticle momentum state in the rest frame
that is boosted to overall 4-velocity $v$ ($v_\mu v^\mu = 1$) by
means of a canonical spin boost $B_c(v)$~\cite{Keister:1991sb}:
\begin{eqnarray}\label{eq:velstat}
\lefteqn{\vert v; \vec{k}_1, \mu_1; \vec{k}_2, \mu_2;\dots;
\vec{k}_n, \mu_n \rangle}\\ &&:= \hat{U}_{B_c(v)} \, \vert
\vec{k}_1, \mu_1; \vec{k}_2, \mu_2;\dots; \vec{k}_n, \mu_n \rangle
\,\, \mathrm{with} \,\, \sum_{i=1}^n \vec{k}_i=0\, .\nonumber
\end{eqnarray}
The $\mu_i$s are the spin projections of the individual particles.
By construction one of the $\vec{k}_i$s is redundant. Velocity
states are orthogonal
\begin{eqnarray}\label{eq:vnorm}
\lefteqn{\!\!\!\!\!\!\!\!\!\langle v^\prime; \vec{k}_1^\prime,
\mu_1^\prime; \vec{k}_2^\prime, \mu_2^\prime;\dots;
\vec{k}_n^\prime, \mu_n^\prime \vert \, v; \vec{k}_1, \mu_1;
\vec{k}_2, \mu_2;\dots; \vec{k}_n, \mu_n \rangle  \nonumber}\\ & &
= v_0 \, \delta^3(\vec{v}^\prime-\vec{v})\, \frac{(2 \pi)^3 2
\omega_{k_n}}{\left( \sum_{i=1}^n \omega_{k_i}\right)^3}
\nonumber\\ && \phantom{=}\times\left( \prod_{i=1}^{n-1} (2\pi)^3
2 \omega_{k_i} \delta^3(\vec{k}_i^\prime-\vec{k}_i)\right) \left(
\prod_{i=1}^{n} \delta_{\mu_i^\prime \mu_i}\right)\,
\end{eqnarray}
and satisfy the completeness relation
\begin{eqnarray}\label{eq:vcompl}
\mathds{1}_{1,2,\dots,n}&=&\sum_{\mu_1=-j_1}^{j_1}
\sum_{\mu_2=-j_2}^{j_2} \dots \sum_{\mu_n=-j_n}^{j_n} \int
\frac{d^3 v}{(2\pi)^3 v_0} \nonumber\\ &&\times \left[
\prod_{i=1}^{n-1}\frac{d^3k_i}{(2 \pi)^3 2 \omega_{k_i}}
\right]\frac{\left(\sum_{i=1}^n \omega_{k_i}\right)^3}{2
\omega_{k_n}}\nonumber\\ & & \times \vert v; \vec{k}_1, \mu_1;
\vec{k}_2, \mu_2;\dots; \vec{k}_n, \mu_n \rangle
\nonumber\\&&\times \langle v; \vec{k}_1, \mu_1; \vec{k}_2,
\mu_2;\dots; \vec{k}_n, \mu_n\vert\, ,
\end{eqnarray}
with $m_i$, $\omega_{k_i}:= \sqrt{m_i^2+\vec{k}_i^2}$, and $j_i$,
being the mass, the energy, and the spin of the $i$th particle,
respectively. Without loss of generality we have taken the $n$th
momentum to be redundant.

\begin{widetext}
One of the big advantages of velocity states as compared with
usual momentum states is their simple behavior under a Lorentz
transformation $\Lambda$:
\begin{eqnarray}\label{eq:vstateboost}
\lefteqn{\hat{U}_\Lambda \vert v; \vec{k}_1, \mu_1; \vec{k}_2,
\mu_2;\dots; \vec{k}_n, \mu_n \rangle} \nonumber\\ &&=
\sum_{\mu_1^\prime, \mu_2^\prime,\dots,\mu_n^\prime}\, \left\{
\prod_{i=1}^n \, D^{j_i}_{\mu_i^\prime \mu_i}\left[R_{\mathrm
W}(v,\Lambda)\right] \right\}
 \vert \Lambda v; \overrightarrow{R_{\mathrm
W}(v,\Lambda)k}_1, \mu_1^\prime; \overrightarrow{R_{\mathrm
W}(v,\Lambda)k}_2, \mu_2^\prime;\dots; \overrightarrow{R_{\mathrm
W}(v,\Lambda)k}_n, \mu_n^\prime \rangle\, ,
\end{eqnarray}
\end{widetext}
with the Wigner-rotation matrix
\begin{equation}\label{eq:wignerrot}
R_{\mathrm W}(v,\Lambda) = B_c^{-1}(\Lambda v)\Lambda B_c(v)\, .
\end{equation}
Since the Wigner rotations are the same for all particles angular
momenta can be added as in non-relativistic quantum mechanics. In
a velocity-state basis the Bakamjian-Thomas type 4-momentum
operator, Eq.~(\ref{eq:momentumop}), is diagonal in the 4-velocity
$v$.

\subsection{\label{subsec:scattering}Electron-meson scattering}
We extract the electromagnetic meson current and the corresponding
form factors from the invariant one-photon-exchange amplitude for
electron-meson scattering. This requires to take the dynamics of
the exchanged photon fully into account. Hence we formulate the
scattering of an electron by a (composite) meson on a Hilbert
space that is a direct sum of $eQ\bar{q}$ and $eQ\bar{q}\gamma$
Hilbert spaces. If the eigenstates $|\psi \rangle$ of the total
mass operator $\hat{M}$ are decomposed into $eQ\bar{q}$ and
$eQ\bar{q}\gamma$ components, i.e. $|\psi \rangle =
|\psi_{eQ\bar{q}} \rangle + |\psi_{eQ\bar{q}\gamma} \rangle$, the
mass-eigenvalue equations for these components may be written in
the form:
\begin{eqnarray}\label{eq:mev}
 \left(\begin{array}{cc} \hat M^{\mathrm{conf}}_{eQ\bar{q}} & \hat K_\gamma \\
 \hat K_\gamma^\dagger & \hat M^{\mathrm{conf}}_{eQ\bar{q}\gamma}\end{array}\right)
 \left(\begin{array}{l}  |\psi_{eQ\bar{q}}\rangle \\
 |\psi_{eQ\bar{q}\gamma}\rangle \end{array}\right)   =
 m \left(\begin{array}{l} |\psi_{eQ\bar{q}}\rangle \\
 |\psi_{eQ\bar{q}\gamma}\rangle \end{array}\right) \, .
\end{eqnarray}
$\hat{K}_\gamma^\dag$ and $\hat{K}_\gamma$ are vertex operators that
describe the emission and absorption of a photon by the electron or
(anti)quark. Without loss of generality we have assumed that the
quark $Q$ ($=c,b$) is the heavy and the antiquark $\bar{q}$
($=\bar{u}, \bar{d}, \bar{s}$) the light mesonic constituent,
respectively. The instantaneous confining interaction between quark
and antiquark is already included in the diagonal elements of this
matrix mass operator, i.e.
\begin{eqnarray}\label{eq:MeC}
\hat{M}^{\mathrm{conf}}_{eQ\bar{q}} =
\hat{M}_{e Q \bar{q}}+ \hat{V}^{(3)}_{\mathrm{conf}}\, ,\nonumber\\
\hat{M}^{\mathrm{conf}}_{eQ\bar{q}\gamma} = \hat{M}_{e Q \bar{q}
\gamma} + \hat{V}^{(4)}_{\mathrm{conf}}\, ,
\end{eqnarray}
with $\hat{V}^{(3)}_{\mathrm{conf}}$ and
$\hat{V}^{(4)}_{\mathrm{conf}}$ denoting the embedding of the
confining $Q\bar{q}$-potential into the 3- and 4-particle Hilbert
spaces~\cite{Keister:1991sb}.

The invariant one-photon exchange amplitude for electron-meson
scattering is now obtained by taking appropriate matrix elements
of the optical potential $\hat{V}_{\mathrm{opt}}(m)$ that enters
the equation for $|\psi_{e Q \bar{q}}\rangle$ after a Feshbach
reduction:
\begin{equation}\label{eq:vopt}
( \hat M^{\mathrm{conf}}_{eQ\bar{q}}+ \underbrace{\hat K_\gamma
(m-\hat M^{\mathrm{conf}}_{eQ\bar{q}\gamma})^{-1}\hat
K_\gamma^\dagger}_{\hat V_{\mathrm{opt}}(m)} )
|\psi_{eQ\bar{q}}\rangle
 = |\psi_{eQ\bar{q}}\rangle\, .
\end{equation}
What we need are matrix elements of the optical potential
$\hat{V}_{\mathrm{opt}}(m)$ between (velocity) eigenstates of the
channel mass operator
\begin{eqnarray}\label{eq:veigenst}
\lefteqn{M^{\mathrm{conf}}_{eQ\bar{q}} \vert\,  \underline{v};
\vec{\underline{k}}_e, \underline{\mu}_e;
\vec{\underline{k}}_\alpha,\underline{\mu}_\alpha, \alpha \rangle}\nonumber\\
 &&\qquad = (\omega_{\underline{k}_e} +
\omega_{\underline{k}_\alpha} ) \vert\, \underline{v};
\vec{\underline{k}}_e, \underline{\mu}_e;
\vec{\underline{k}}_\alpha,\underline{\mu}_\alpha,  \alpha \rangle
\, . \\ && \nonumber
\end{eqnarray}
$\underline{\mu}_\alpha$ denotes the spin orientation of the
confined $q\bar{q}$ bound state, $\alpha$ is a shorthand notation
for the remaining discrete quantum numbers necessary to specify it
uniquely. The energy of the $Q\bar{q}$ bound state with quantum
numbers $\alpha$ and mass $m_\alpha$ is
$\omega_{\underline{k}_\alpha}=(m_\alpha^2+
\vec{\underline{k}}_\alpha^2)^{1/2}$. Eigenstates of
$M^{\mathrm{conf}}_{eQ\bar{q}\gamma}$ are introduced in an
analogous way. Later on we will also need (velocity) eigenstates
of the free mass operators $M_{eQ\bar{q}}$ and
$M_{eQ\bar{q}\gamma}$. To make a clear distinction between states
with a confined $Q\bar{q}$ pair from those with a free $Q\bar{q}$
pair we underline velocities, momenta and spin projections for the
former. For the calculation of electromagnetic meson form factors
only on-shell matrix elements of $\hat{V}_{\mathrm{opt}}(m)$  are
required with the discrete quantum numbers $\alpha$ being those of
the meson of interest. The further analysis of these on-shell
matrix elements is accomplished by inserting completeness
relations for the eigenstates of
$M^{\mathrm{conf}}_{eQ\bar{q}\gamma}$, $M_{eQ\bar{q}}$ and
$M_{eQ\bar{q}\gamma}$ at the appropriate places:
\begin{eqnarray}\label{eq:voptclust}
\lefteqn{\langle  \underline{v}^\prime;
\vec{\underline{k}}_e^\prime, \underline{\mu}_e^\prime;
\vec{\underline{k}}_\alpha^\prime,\underline{\mu}_\alpha^\prime,
\alpha \vert\,
\hat{V}_{\mathrm{opt}}(m)
\vert\, \underline{v}; \vec{\underline{k}}_e, \underline{\mu}_e;
\vec{\underline{k}}_\alpha,\underline{\mu}_\alpha,
\alpha \rangle_{\mathrm{os}}} \nonumber \\
&&= \langle  \underline{v}^\prime; \vec{\underline{k}}_e^\prime,
\underline{\mu}_e^\prime;
\vec{\underline{k}}_\alpha^\prime,\underline{\mu}_\alpha^\prime,
\alpha \vert\, \mathds{1}_{e Q \bar{q}}\,\hat{K}_{\gamma}
\mathds{1}_{e Q \bar{q} \gamma}\left(\hat{M}_{e Q\bar q \gamma}^{\text{conf}}
\!-\! m\right)^{-1} \nonumber\\
& &\quad\times \mathds{1}_{e Q \bar{q} \gamma}^{\mathrm{conf}}
\mathds{1}_{e Q \bar{q} \gamma} \, \hat{K}^\dag_{\gamma}
\mathds{1}_{e Q \bar{q}}\, \vert\, \underline{v};
\vec{\underline{k}}_e, \underline{\mu}_e;
\vec{\underline{k}}_\alpha,\underline{\mu}_\alpha,  \alpha
\rangle_{\mathrm{os}}\,
. \nonumber\\
\end{eqnarray}
\lq\lq os\rq\rq\ means on-shell, i.e.
$m=\omega_{\underline{k}_e}+\omega_{\underline{k}_\alpha} =
\omega_{\underline{k}_e^\prime}+\omega_{\underline{k}_\alpha^\prime}$,
$\omega_{\underline{k}_e}=\omega_{\underline{k}_e^\prime}$ and
$\omega_{\underline{k}_\alpha}=\omega_{\underline{k}_\alpha^\prime}$.
After insertion of the completeness relations one ends up with
matrix elements of the form
$\langle v; \vec{k}_e, \mu_e; \vec{k}_Q, \mu_Q; \vec{k}_{\bar{q}},
\mu_{\bar{q}} \vert\, \underline{v}; \vec{\underline{k}}_e,
\underline{\mu}_e;
\vec{\underline{k}}_\alpha,\underline{\mu}_\alpha, \alpha
\rangle$,
$\langle v; \vec{k}_e, \mu_e; \vec{k}_Q, \mu_Q; \vec{k}_{\bar{q}},
\mu_{\bar{q}};\vec{k}_{\gamma}, \mu_{\gamma} \vert\,
\underline{v}; \vec{\underline{k}}_e, \underline{\mu}_e;
\vec{\underline{k}}_\alpha, \underline{\mu}_\alpha, \alpha;
\vec{\underline{k}}_\gamma, \underline{\mu}_\gamma\rangle$,
$ \langle v^\prime; \vec{k}_e^\prime,\! \mu_e^\prime;
\vec{k}_Q^\prime,\! \mu_Q^\prime; \vec{k}_{\bar{q}}^\prime,\!
\mu_{\bar{q}}^\prime; \vec{k}_\gamma^\prime,\! \mu_\gamma^\prime
\vert \,\hat{K}^\dag\, \vert v\,  ; \vec{k}_e,\! \mu_e; \vec{k}_Q,\!
\mu_Q; \vec{k}_{\bar{q}},\! \mu_{\bar{q}} \rangle $
and the Hermitian conjugates, respectively. The first two are just
wave functions of the confined $Q\bar{q}$ pair and a free electron
(and photon). The third describes the transition from a free
$Q\bar{q}e$ state to a free $Q\bar{q}e\gamma$ state by emission of a
photon and is calculated from the usual interaction density
${\mathcal{L}}^{\mathrm{em}}_{\mathrm{int}}(x)$ of spinor quantum
electrodynamics~\cite{Klink:2000pp}:
\begin{eqnarray}\label{eq:emvertex}\lefteqn{\hspace{-0.9cm}
\langle v^\prime; \vec{k}_e^\prime,\! \mu_e^\prime;
\vec{k}_Q^\prime,\! \mu_Q^\prime; \vec{k}_{\bar{q}}^\prime,\!
\mu_{\bar{q}}^\prime; \vec{k}_\gamma^\prime,\! \mu_\gamma^\prime
\vert \,\hat{K}^\dag\, \vert v\,  ; \vec{k}_e,\! \mu_e; \vec{k}_Q,\!
\mu_Q; \vec{k}_{\bar{q}},\! \mu_{\bar{q}} \rangle} \nonumber\\
&=& N v_0 \delta^3(\vec{v}^\prime - \vec{v})\, \langle
\vec{k}_e^\prime,\! \mu_e^\prime; \vec{k}_Q^\prime,\! \mu_Q^\prime;
\vec{k}_{\bar{q}}^\prime,\! \mu_{\bar{q}}^\prime;
\vec{k}_\gamma^\prime,\! \mu_\gamma^\prime \vert \nonumber\\
&&\times \hat{\mathcal{L}}^{\mathrm{em}}_{\mathrm{int}}(0)\, \vert
\vec{k}_e,\! \mu_e; \vec{k}_Q,\! \mu_Q; \vec{k}_{\bar{q}},\!
\mu_{\bar{q}} \rangle\, .
\end{eqnarray}
The normalization factor $N$ is determined by the normalization of
the velocity states. Explicit expressions for all the matrix
elements are given in Ref.~\cite{Biernat:2009my}. Using these
analytical results we can show that the on-shell matrix elements of
the optical potential have the structure that one expects from the
invariant one-photon-exchange amplitude, namely the electron current
$j_e^\mu$ contracted with the hadron current $\tilde{J}^\nu_{[\alpha]}$ and
multiplied with the covariant photon propagator $(-g_{\mu\nu}/Q^2)$
(times some kinematical factors):
\begin{widetext}
\begin{eqnarray}\label{eq:voptos}
\lefteqn{\langle  \underline{v}^\prime;
\vec{\underline{k}}_e^\prime, \underline{\mu}_e^\prime;
\vec{\underline{k}}_\alpha^\prime,\underline{\mu}_\alpha^\prime,
\alpha \vert\,
\hat{V}_{\mathrm{opt}}(m)
\vert\, \underline{v}; \vec{\underline{k}}_e, \underline{\mu}_e;
\vec{\underline{k}}_\alpha,\underline{\mu}_\alpha^\prime, \alpha \rangle_{\mathrm{os}}}
\nonumber \\
&&=\underline{v}_0 \delta^3 (\vec{\underline{v}}^{\, \prime} -
\vec{\underline{v}}\, )\, \frac{(2 \pi)^3
}{\sqrt{(\omega_{\underline{k}_e^{\prime}}+
\omega_{\underline{k}_\alpha^{\prime}})^3}
\sqrt{(\omega_{\underline{k}_e^{\phantom{\prime}}}+
\omega_{\underline{k}_\alpha^{\phantom{\prime}}})^3}}
(-e^2)\underbrace{\,\bar{u}_{\underline{\mu}_e^\prime}
(\vec{\underline{k}}_e^\prime)\gamma^\mu u_{\underline{\mu}_e}
(\vec{\underline{k}}_e)}_{j_e^\mu(\vec{\underline{k}}_e^\prime,
\underline{\mu}_e^\prime;\vec{\underline{k}}_e,
\underline{\mu}_e)}\frac{(-g_{\mu \nu})}{Q^2}\underbrace{ (Q_Q
J_Q^\nu(\dots) + Q_{\bar{q}}J_{\bar
q}^\nu(\dots))}_{\tilde{J}_{[\alpha]}^\nu(\vec{\underline{k}}_\alpha^\prime,
\underline{\mu}_\alpha^\prime;\vec{\underline{k}}_\alpha,
\underline{\mu}_\alpha)}\, .
\end{eqnarray}
Here we have introduced the (negative) square of the (space-like)
4-momentum-transfer $Q^2=-\underline{q}_\mu \underline{q}^\mu$, with
$\underline{q}^\mu = (\underline{k}_\alpha-
\underline{k}_\alpha^\prime)^\mu = (\underline{k}_e^\prime -
\underline{k}_e)^\mu$. $e$, $Q_Q |\,e|$ and $Q_{\bar{q}} |\,e|$
denote the electric charges of the electron, the quark and the
antiquark, respectively. We want to emphasize that the kinematical
factor in front of Eq.~(\ref{eq:voptos}) and thus the normalization
of the meson current $\tilde{J}^\nu_{[\alpha]}$ is uniquely fixed.
It must be identical with the one that comes out if the optical
potential is derived in an analogous way for the scattering of an
electron by a point-like meson with discrete quantum numbers
$\alpha$ (see Refs.~\cite{Biernat:2009my,Biernat:2011mp}). Since the
point-like current is known this kinematical factor can be uniquely
identified. The two parts of the meson current, $J_Q^\nu$ and
$J_{\bar q}^\nu$, correspond to the coupling of the photon to the
quark or the antiquark, respectively. If $\alpha$ are the discrete
quantum numbers of a pseudoscalar ground-state meson
($\underline{\mu}_\alpha=\underline{\mu}_\alpha^\prime=0$) it has to
be a pure s-wave and  we find that
\begin{eqnarray}\label{eq:JQ}
&& J^{\nu}_Q
(\vec{\underline{k}}_\alpha^\prime,\vec{\underline{k}}_\alpha)=
\frac{\sqrt{\omega_{\underline{k}_\alpha}\omega_{\underline{k}_\alpha^{\prime}}}}{4
\pi} \int\, \frac{d^3\tilde{k}_{\bar{q}}^\prime}{2 \omega_{k_Q}}\,
\sqrt{\frac{\omega_{{k}_Q}+\omega_{{k}_{\bar{q}}}}
{\omega_{{k}^\prime_Q}+\omega_{{k}^\prime_{\bar{q}}}}} \,
\sqrt{\frac{\omega_{\tilde{k}^\prime_Q}+\omega_{\tilde{k}^\prime_{\bar{q}}}}
{\omega_{\tilde{k}_Q}+\omega_{\tilde{k}_{\bar{q}}}}} \,
\sqrt{\frac{\omega_{\tilde{k}_Q} \omega_{\tilde{k}_{\bar{q}}}}
{\omega_{\tilde{k}^\prime_Q} \omega_{\tilde{k}^\prime_{\bar{q}}}}}
 \,  \bigg\{\!\sum_{\mu_Q,\mu_Q^\prime
=\pm \frac{1}{2}}\!\!\!
\bar{u}_{\mu_Q^\prime}(\vec{k}_Q^\prime)\,\gamma^\nu\,
u_{\mu_Q}(\vec{k}_Q)  \\ &&\times D^{1/2}_{\mu_Q\mu_Q^\prime}\!
\left[ \!R_{\mathrm{W}}\!\left(\frac{\tilde{k}_{Q}}{m_Q},
B_c(v_{Q\bar{q}})\right)\,
\!R^{-1}_{\mathrm{W}}\!\left(\frac{\tilde{k}_{\bar{q}}}{m_{\bar{q}}},
B_c(v_{Q\bar{q}})\right)\,
\!R_{\mathrm{W}}\!\left(\frac{\tilde{k}_{\bar q}^\prime}{m_{\bar
q}}, B_c(v_{Q\bar{q}}^\prime)\right)\,
\!R^{-1}_{\mathrm{W}}\!\left(\frac{\tilde{k}_{Q}^\prime}{m_{Q}},
B_c(v_{Q\bar{q}}^\prime)\right)\right] \bigg\}\, \psi^\ast\,(\vert
\vec{\tilde{k}}_{\bar{q}}^\prime\vert)\,  \psi
 \,(\vert \vec{\tilde{k}}_{\bar{q}}\vert)\,\, . \nonumber
\end{eqnarray}
\end{widetext}
The corresponding expression for $J^{\nu}_{\bar q}$ is obtained by
interchanging $Q$ and $\bar q$ in Eq.~(\ref{eq:JQ}). The quantities
with a tilde are defined in the rest frame of the $Q\bar{q}$
subsystem. The s-wave bound-state wave function $\psi(\kappa)$ is
also defined in this frame and normalized according to
\begin{equation}\label{eq:psinorm}
\int_0^\infty \, d\kappa \kappa^2\, \psi^\ast (\kappa)\,
\psi(\kappa) = 1 \, .
\end{equation}
The transformation between the $Q\bar{q}$ rest frame and the
$Q\bar{q}e$ rest frame is accomplished by means of a canonical spin
boost~\cite{Keister:1991sb}
\begin{equation}\label{canonical boost}
B_c(v) = \left( \begin{array}{cc} v^0 & \mathbf v^T \\ \mathbf v &
\mathbf 1 +\frac{v^0 -1}{\mathbf v ^2 }\mathbf v \mathbf v ^T
\end{array}\right)
\end{equation}
with
\begin{equation}
v=v_{Q\bar{q}}=\frac{k_Q + k_{\bar{q}}}{m_{Q\bar{q}}}
\end{equation}
and \begin{equation}
m_{Q\bar{q}}=\omega_{\tilde{k}_Q}+\omega_{\tilde{k}_{\bar{q}}} =
\sqrt{(\omega_{k_Q}+\omega_{k_{\bar{q}}})^2-(\vec{k}_Q+\vec{k}_{\bar{q}})^2}\,
\end{equation}
denoting the invariant mass of the (unbound) $Q\bar{q}$ pair. Here
it is useful to note that, due to our center-of-mass kinematics,
$\vec{k}_e+\vec{k}_Q+\vec{k}_{\bar{q}}=\vec{k}_e+\vec{\underline{k}}_\alpha=0$
and hence $\vec{k}_Q+\vec{k}_{\bar{q}}=\vec{\underline{k}}_\alpha$
such that
$\vec{v}_{Q\bar{q}}=\vec{\underline{k}}_\alpha/m_{Q\bar{q}}$.
Analogous relations hold for $v_{Q\bar{q}}^\prime$ and the primed
momenta. This implies further that not all of the 4-momentum that is
transferred via the photon to the $Q\bar{q}$ bound state is also
transferred to the active constituent. Only the 3-momentum transfer
is the same. For the quark being the active particle we have, e.g.,
$\vec{\underline{q}}=\vec{\underline{k}}_\alpha-
\vec{\underline{k}}_\alpha^\prime=
\vec{k}_Q-\vec{k}_Q^\prime=:\vec{q}_\mathrm{quark}$. On the other
hand one has
$\omega_{\underline{k}_\alpha}=\omega_{\underline{k}_\alpha^\prime}$
and hence $\underline{q}^0=0$, whereas, in general,
$q^0_\mathrm{quark}:=\omega_{k_Q}-\omega_{k_Q^\prime}\neq 0$. If the
photon couples to the quark, the spectator condition $k_{\bar q}=k_{\bar q}^\prime$
for the antiquark implies the relation:
\begin{eqnarray}\label{eq:kktildep}
\tilde{k}_{\bar q} &=& B_c^{-1}(v_{Q\bar{q}})\, k_{\bar q}
= B_c^{-1}(v_{Q\bar{q}})\, k_{\bar q}^\prime\nonumber \\
&=& B_c^{-1}(v_{Q\bar{q}})\, B_c(v_{Q\bar{q}}^\prime)\tilde{k}_{\bar
q}^\prime \, .
\end{eqnarray}
The 4-momenta $\tilde{k}_Q^{(\prime)}$ for the active quark are then
uniquely determined by $\vec{\tilde k}_{Q}^{(\prime)}=-\vec{\tilde
k}_{\bar q}^{(\prime)}$. Associated with the boosts that connect
incoming and outgoing wave functions are Wigner rotations of the
quark and antiquark spins. The corresponding Wigner $D$ functions
can be combined to the single one showing up in Eq.~(\ref{eq:JQ}) by
means of the spectator conditions and the Clebsch coefficients that
couple the quark and the antiquark spins to zero meson spin (see,
e.g., Ref.~\cite{Biernat:2011mp}).

Having obtained the microscopic expression for the
electromagnetic meson current $\tilde{J}^\nu_{[\alpha]}$ (cf. Eqs.
(\ref{eq:voptos}) and (\ref{eq:JQ})), we will show in the sequel how
the derivation of the weak current, as occurring in semileptonic
meson decays, is accomplished within our relativistic
coupled-channel framework.

\subsection{\label{subsec:decay}Semileptonic meson decay}
\begin{figure}[t!]
\includegraphics[width=0.45\textwidth]{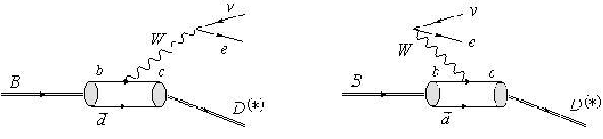}
\caption{The two time ordered contributions to the semileptonic weak
decay of a $\bar{B}^0$ into a $D^{(\ast)+}$ meson.} \label{fig:decay}\end{figure}

In order to get the full (leading-order) invariant amplitude for the
semileptonic weak decay of a heavy-light meson $\alpha$ into another
heavy-light meson $\alpha^\prime$ one needs at least 4 channels.
This can be seen immediately, if one decomposes this amplitude into
its time-ordered contributions. This decomposition is depicted in
Fig.~\ref{fig:decay} for the $\bar{B}^0\rightarrow D^{(\ast)+} e
\bar{\nu}_e$ decay on which we will concentrate in the following. In
addition to the incoming $b\bar{d}$ channel and the outgoing
$c\bar{d}e\bar{\nu}_e$ channel one needs a $c\bar{d}W$ and a
$b\bar{d}We\bar{\nu}_e$ channel to account for the intermediate
states in which the $W$-boson is in flight. The matrix mass operator
acting on all these channels has the form
\begin{widetext}
\begin{equation}\label{eq:massopdecay}
 \left(\begin{array}{cccc} \hat M^{\mathrm{conf}}_{b\bar{d}} & 0 &
 \hat{K}_{c\bar{d}W\rightarrow b\bar{d}} &
 \hat{K}_{b\bar{d}We\bar{\nu}_e\rightarrow b\bar{d}}\\
0 & \hat M^{\mathrm{conf}}_{c\bar{d}e\bar{\nu}_e} &
\hat{K}_{c\bar{d}W\rightarrow c\bar{d}e\bar{\nu}_e} &
\hat{K}_{b\bar{d}We\bar{\nu}_e\rightarrow c\bar{d}e\bar{\nu}_e} \\
\hat{K}^\dag_{c\bar{d}W\rightarrow b\bar{d}} &
\hat{K}^\dag_{c\bar{d}W\rightarrow c\bar{d}e\bar{\nu}_e} & \hat
M^{\mathrm{conf}}_{c\bar{d}W} &
0 \\
\hat{K}^\dag_{b\bar{d}We\bar{\nu}_e\rightarrow b\bar{d}} &
\hat{K}^\dag_{b\bar{d}We\bar{\nu}_e\rightarrow c\bar{d}e\bar{\nu}_e}
& 0 & \hat M^{\mathrm{conf}}_{b\bar{d}We\bar{\nu}_e}
\end{array}\right)\, .
\end{equation}
As in the electromagnetic case an instantaneous confining potential
between the quark-antiquark pair is included in the channel mass
operators on the diagonal. What we are interested in is the
transition from the $b\bar{d}$ to the $c\bar{d}e\bar{\nu}_e$
channel. As can be seen from Eq.~(\ref{eq:massopdecay}) this cannot
happen directly. It only works via the intermediate states that
contain the $W$. The corresponding (optical) transition potential
$\hat V_{\mathrm{opt}}^{b\bar{d}\rightarrow
c\bar{d}e\bar{\nu}_e}(m)$ may be again obtained by applying a
Feshbach reduction to eliminate the $c\bar{d}W$ and the
$b\bar{d}We\bar{\nu}_e$ channels such that one ends up with a mass
eigenvalue problem for the (coupled) $b\bar{d}$ and
$c\bar{d}e\bar{\nu}_e$ system. The transition potential has then the
form
\begin{equation}
\hat V_{\mathrm{opt}}^{b\bar{d}\rightarrow c\bar{d}e\bar{\nu}_e}(m)
= \hat{K}_{c\bar{d}W\rightarrow c\bar{d}e\bar{\nu}_e}
(m-M^{\mathrm{conf}}_{c\bar{d}W})^{-1}\hat{K}^\dag_{c\bar{d}W\rightarrow
b\bar{d}} + \hat{K}_{b\bar{d}We\bar{\nu}_e\rightarrow
c\bar{d}e\bar{\nu}_e} (m-\hat
M^{\mathrm{conf}}_{b\bar{d}We\bar{\nu}_e})^{-1}
 \hat{K}^\dag_{b\bar{d}We\bar{\nu}_e\rightarrow b\bar{d}}\, .
\end{equation}
\end{widetext}
The two terms on the right-hand side correspond to the two
time-orderings of the $W$ exchange that are depicted in
Fig.~\ref{fig:decay}.

Like in the electromagnetic case the weak hadronic current and the
$B\rightarrow D^{(\ast)}$ decay form factors are extracted from
on-shell matrix elements of $\hat
V_{\mathrm{opt}}^{b\bar{d}\rightarrow c\bar{d}e\bar{\nu}_e}(m)$,
i.e. from
\begin{equation}\label{eq:vopttrans}
\langle \underline{v}^\prime; \vec{\underline{k}}_e^\prime,
\underline{\mu}_e^\prime; \vec{\underline{k}}_{\bar{\nu}_e}^\prime;
\vec{\underline{k}}_{\alpha^\prime}^\prime,\underline{\mu}_{\alpha^\prime}^\prime,
\alpha^\prime \vert \hat V_{\mathrm{opt}}^{b\bar{d}\rightarrow
c\bar{d}e\bar{\nu}_e}(m)\vert \vec{\underline{k}}_{\alpha},
\underline{\mu}_{\alpha}, \alpha \rangle_{\mathrm{os}}\, ,
\end{equation}
where the discrete quantum numbers $\alpha$ and $\alpha^{\prime}$ of
the confined heavy-light system are those of the $B$ and
$D^{(\ast)}$, respectively. \lq\lq On shell\rq\rq means now that
$m=m_B = \omega_{\underline{k}_\alpha} =
\omega_{\underline{k}_{\alpha^\prime}^\prime}
+\omega_{\underline{k}_e^\prime}+\omega_{\underline{k}_{\bar{\nu}_e}^\prime}$.
For the analysis of these matrix elements we can proceed as in the
electromagnetic case. One has to insert the appropriate completeness
relations at the pertinent places. This leads again to wave
functions for the confined $Q^{(\prime)}\bar{q}$ pair in combination
with a free $W$ and/or a free $e$-$\bar{\nu}_e$ pair. The matrix
elements of the weak vertex operators $\hat{K}_{c\bar{d}W\rightarrow
b\bar{d}}$, etc., can be derived from the weak interaction density
$\mathcal{L}^{\mathrm{wk}}_{\mathrm{int}}(x)$ in analogy to
Eq.~(\ref{eq:emvertex}). After insertion of the analytical
expressions for the wave functions and the vertex matrix elements
into Eq.~(\ref{eq:vopttrans}) we observe again that the on-shell
matrix elements of $V_{\mathrm{opt}}^{b\bar{d}\rightarrow
c\bar{d}e\bar{\nu}_e}(m)$ have the same structure as the invariant
$B\rightarrow D^{(\ast)} e \bar{\nu}_e$ decay amplitude that results
from leading-order covariant perturbation theory:
\begin{eqnarray}\label{eq:voptcov}
\lefteqn{\langle \underline{v}^\prime; \vec{\underline{k}}_e^\prime,
\underline{\mu}_e^\prime; \vec{\underline{k}}_{\bar{\nu}_e}^\prime;
\vec{\underline{k}}_{\alpha^\prime}^\prime,\underline{\mu}_{\alpha^\prime}^\prime,
\alpha^\prime \vert \hat V_{\mathrm{opt}}^{b\bar{d}\rightarrow
c\bar{d}e\bar{\nu}_e}(m)\vert \vec{\underline{k}}_{\alpha},
\underline{\mu}_{\alpha}, \alpha \rangle_{\mathrm{os}}}\nonumber\\
&=& \underline{v}_0 \delta^3 (\vec{\underline{v}}^{\, \prime} -
\vec{\underline{v}}\, )\, \frac{(2 \pi)^3
}{\sqrt{(\omega_{\underline{k}_e^{\prime}}+\omega_{\underline{k}_{\bar{\nu}_e}^{\prime}}
+\omega_{\underline{k}_{\alpha^\prime}^{\prime}})^3} \sqrt{
\omega_{\underline{k}_\alpha^{\phantom{\prime}}}^3}}\nonumber\\&&\times
\frac{e^2}{2\sin^2\vartheta_{\mathrm{w}}}V_{cb}\,
\frac{1}{2}\underbrace{\bar{u}_{\underline{\mu}_e^\prime}
(\vec{\underline{k}}_e^\prime)\gamma^\mu (1-\gamma^5)
v_{\underline{\mu}_{\bar{\nu}_e}^\prime}
(\vec{\underline{k}}_{\bar{\nu}_e}^\prime) }_{j_{\nu_e\rightarrow
e}^\mu(\vec{\underline{k}}_e^\prime,
\underline{\mu}_e^\prime;\vec{\underline{k}}_{\bar{\nu}_e}^\prime,
\underline{\mu}_{\bar{\nu}_e}^\prime)}\\&&\times\frac{(-g_{\mu
\nu})}{(\underline{k}^\prime_e+\underline{k}^\prime_{\bar{\nu}_e})^2-m_W^2}
\frac{1}{2}{J_{\alpha\rightarrow
\alpha^\prime}^\nu(\vec{\underline{k}}_{\alpha^\prime}^\prime,
\underline{\mu}_{\alpha^\prime}^\prime;\vec{\underline{k}}_\alpha,
\underline{\mu}_\alpha)}\, .\nonumber
\end{eqnarray}

Here $\vartheta_{\mathrm{w}}$ denotes the electroweak mixing angle
and $e$ the usual elementary electric charge and $V_{cb}$ is the CKM
matrix element occurring at the $Wbc$-vertex. Like in the
electromagnetic case the kinematical factor in front and hence the
normalization of the weak hadronic transition current
$J_{\alpha\rightarrow \alpha^\prime}^\nu$ is uniquely fixed. The
only difference between the two time orderings contributing to the
decay amplitude comes from the propagator in the intermediate state.
Summing the two propagators (and dividing by
$2\omega_{\underline{k}_W}$) leads to the covariant $W$ propagator
that occurs in Eq.~(\ref{eq:voptcov}).

Let now $\alpha$ be the quantum numbers of a $B$ meson and
$\alpha^\prime$ those of a $D$ meson. Since $B$ and $D$ have to be
pure s-wave the weak transition current becomes
\begin{widetext}
\begin{eqnarray}\label{eq:Jwkpsps}
 J^{\nu}_{B\rightarrow D}
(\vec{\underline{k}}_D^\prime;\vec{\underline{k}}_B=\vec{0})&=&
\frac{\sqrt{\omega_{\underline{k}_B}\omega_{\underline{k}_D^{\prime}}}}{4
\pi} \int\, \frac{d^3\tilde{k}_{\bar{q}}^\prime}{2 \omega_{k_b}}\,
\sqrt{\frac{\omega_{\tilde{k}^\prime_c}+\omega_{\tilde{k}^\prime_{\bar{q}}}}
{\omega_{{k}^\prime_c}+\omega_{{k}^\prime_{\bar{q}}}}} \,
\sqrt{\frac{\omega_{\tilde{k}_b} \omega_{\tilde{k}_{\bar{q}}}}
{\omega_{\tilde{k}^\prime_c} \omega_{\tilde{k}^\prime_{\bar{q}}}}}
 \,  \bigg\{\!\sum_{\mu_b,\mu_c^\prime
=\pm \frac{1}{2}}\!\!\!
\bar{q}_{\mu_c^\prime}(\vec{k}_c^\prime)\,\gamma^\nu\,(1-\gamma^5)\,
u_{\mu_b}(\vec{k}_b)\nonumber  \\ &&\times
D^{1/2}_{\mu_b\mu_c^\prime}\!
\left[\!R_{\mathrm{W}}\!\left(\frac{\tilde{k}_{\bar
q}^\prime}{m_{\bar q}}, B_c(v_{c\bar{q}}^\prime)\right)\,
\!R^{-1}_{\mathrm{W}}\!\left(\frac{\tilde{k}_{c}^\prime}{m_{c}},
B_c(v_{c\bar{q}}^\prime) \right)\right] \bigg\}\,
\psi^\ast_{D}\,(\vert \vec{\tilde{k}}_{\bar{q}}^\prime\vert)\, \psi_B
 \,(\vert \vec{\tilde{k}}_{\bar{q}}\vert)\, .
\end{eqnarray}
$\psi_B$ as well as $\psi_D$ (and in the following $\psi_{D^\ast}$)
are normalized like in Eq.~(\ref{eq:psinorm}). The primed
constituents' momenta are related by
${k}_c^\prime=B_c(v^\prime_{c\bar{q}})\tilde{k}_c^\prime$,
${k}_{\bar{q}}^\prime=B_c(v^\prime_{c\bar{q}})\tilde{k}_{\bar{q}}^\prime$,
where $\vec{\tilde{k}}_{\bar{q}}^\prime=-\vec{\tilde{k}}_c^\prime$
and
$\vec{v}^\prime_{c\bar{q}}=\vec{\underline{k}}_D^\prime/(\omega_{\tilde{k}^\prime_c}+
\omega_{\tilde{k}^\prime_{\bar{q}}})$. Since the $B$ meson is at
rest and the antiquark obeys a spectator condition the unprimed
momenta are then given by
$\vec{k}_{\bar{q}}=\vec{\tilde{k}}_{\bar{q}}=
-\vec{\tilde{k}}_{b}=-\vec{k}_{b}=\vec{k}_{\bar{q}}^\prime$.

If $\alpha^\prime$ are the quantum numbers of a $D^\ast$ meson one
has a pseudoscalar to vector transition. For such a transition both,
the vector and axial-vector part of the weak current contribute. For
$B$ and $D^\ast$ being again pure s-wave (neglecting possible d-wave
contributions in $D^\ast$) the weak transition current differs then
from the one in Eq.~(\ref{eq:Jwkpsps}) mainly by Wigner $D$
functions and Clebsch Gordans:
\begin{eqnarray}\label{eq:Jwkpsv}
 J^{\nu}_{B\rightarrow D^\ast}
(\vec{\underline{k}}_{D^\ast}^\prime,\underline{\mu}_{D^\ast}^\prime;\vec{\underline{k}}_B=\vec{0})&=&
\!\!
\frac{\sqrt{\omega_{\underline{k}_B}\omega_{\underline{k}_{D^\ast}^{\prime}}}}{4
\pi} \int\, \frac{d^3\tilde{k}_{\bar{q}}^\prime}{2 \omega_{k_b}}\,
\sqrt{\frac{\omega_{\tilde{k}^\prime_c}+\omega_{\tilde{k}^\prime_{\bar{q}}}}
{\omega_{{k}^\prime_c}+\omega_{{k}^\prime_{\bar{q}}}}} \,
\sqrt{\frac{\omega_{\tilde{k}_b} \omega_{\tilde{k}_{\bar{q}}}}
{\omega_{\tilde{k}^\prime_c} \omega_{\tilde{k}^\prime_{\bar{q}}}}}
 \,
 \bigg\{\!\sum_{\mu_b,\mu_c^\prime,\tilde\mu_c^\prime,\tilde\mu_{\bar
 q}^\prime=\pm \frac{1}{2}}
 \!\!\!\!\!\!
\bar{q}_{\mu_c^\prime}(\vec{k}_c^\prime)\,\gamma^\nu\,
(1-\gamma^5) u_{\mu_b}(\vec{k}_b)  \nonumber \\
&&\hspace{-4cm}\times \sqrt{2} (-1)^{\frac{1}{2}-\mu_b}
C^{1\mu^\prime_{\!D^\ast}}_{\frac{1}{2}\tilde\mu_c^\prime\frac{1}{2}
\tilde{\mu}_{\bar{q}}^\prime}\, D^{1/2}_{\tilde\mu_c^\prime
\mu_c^\prime}\!
\left[\!R^{-1}_{\mathrm{W}}\!\left(\frac{\tilde{k}_{c}^\prime}{m_{c}},
B_c(v_{c\bar{q}}^\prime)\right)\right]\,D^{1/2}_{\tilde\mu_{\bar{q}}^\prime
-\mu_b}\!\left[ \!R^{-1}
_{\mathrm{W}}\!\left(\frac{\tilde{k}^\prime_{\bar q}}{m_{\bar q}},
B^{-1}_c(v_{c\bar{q}}^\prime)\right)\right] \bigg\}\,
\psi^\ast_{D^{\ast}}\,(\vert \vec{\tilde{k}}_{\bar{q}}^\prime\vert)\, \psi_B
 \,(\vert \vec{\tilde{k}}_{\bar{q}}\vert)\, .
\end{eqnarray}
\end{widetext}
The next step will be to analyze the covariant structure of the
microscopic meson (transition) currents (\ref{eq:JQ}),
(\ref{eq:Jwkpsps}) and (\ref{eq:Jwkpsv}) and to identify the electromagnetic and weak
form factors.

\section{\label{sec:currents}Currents and form factors}

\subsection{\label{subsec:emff}Electromagnetic form factor}
Before we are going to extract the electromagnetic form factor for a
pseudoscalar heavy-light meson we notice that the electromagnetic
current
$\tilde{J}_{[\alpha]}^\nu(\vec{\underline{k}}_\alpha^\prime;\vec{\underline{k}}_\alpha)$
which we have derived in Eqs. (\ref{eq:voptos}) and (\ref{eq:JQ})
still does not transform appropriately under Lorentz
transformations. Since we are using velocity states,
$\vec{\underline{k}}_\alpha$ and $\vec{\underline{k}}_\alpha^\prime$
are momenta defined in the center of mass of the electron-meson
system. As a consequence
$\tilde{J}_{[\alpha]}^\nu(\vec{\underline{k}}_\alpha^\prime;\vec{\underline{k}}_\alpha)$
does not behave like a 4-vector under a Lorentz transformation
$\Lambda$. It rather transforms by the Wigner rotation
$R_W(v,\Lambda)$. Going, however, back to the physical meson momenta
$\underline{p}_{\alpha}^{(\prime)}=B_c(v)
\underline{k}_\alpha^{(\prime)}$ gives a current with the desired
transformation properties~\cite{Biernat:2009my,Biernat:2011mp}:
\begin{equation}
\tilde{J}_{[\alpha]}^\nu(\vec{\underline{p}}_\alpha^\prime;\vec{\underline{p}}_\alpha):=
[B_c(\underline{v})]^\nu_{\phantom{\nu}\rho}
\tilde{J}_{[\alpha]}^\rho(\vec{\underline{k}}_\alpha^\prime;\vec{\underline{k}}_\alpha)\,
.
\end{equation}
$\tilde{J}_{[\alpha]}^\nu(\vec{\underline{p}}_\alpha^\prime;\vec{\underline{p}}_\alpha)$
transforms like a 4-vector and is a conserved current, i.e.
$(\underline{p}_\alpha-\underline{p}_\alpha^\prime)_\nu
\tilde{J}_{[\alpha]}^\nu(\vec{\underline{p}}_\alpha^\prime;\vec{\underline{p}}_\alpha)=0$
~\cite{Biernat:2009my,Biernat:2011mp}. If it would be a perfect
model for the electromagnetic current of a pseudoscalar heavy-light
meson it should be possible to write it in the form
\begin{equation}
J_{[\alpha]}^\nu(\vec{\underline{p}}_\alpha^\prime;\vec{\underline{p}}_\alpha)
=  (\underline{p}_\alpha+\underline{p}^\prime_\alpha)^\nu \,
F(Q^2)
\end{equation}
for arbitrary values of $\underline{p}_\alpha$ and
$\underline{p}^\prime_\alpha$. This, however, does not hold in our
case. The reason is that our derivation of the current makes use of
the Bakamjian-Thomas construction which guarantees Poincar\'e
invariance, but is known to cause problems with cluster
separability~\cite{Keister:1991sb}. As a consequence of wrong
cluster properties the hadronic current we get may also depend on
the electron momenta. We find indeed that
$\tilde{J}_{[\alpha]}^\nu(\vec{\underline{p}}_\alpha^\prime;\vec{\underline{p}}_\alpha)$
cannot be expressed in terms of hadronic covariants only, but one
needs one additional (current conserving) covariant, which is the
sum of incoming and outgoing electron momenta:
\begin{eqnarray}\label{eq:Jemcovdec}
\tilde{J}_{[\alpha]}^\nu(\vec{\underline{p}}_\alpha^\prime;\vec{\underline{p}}_\alpha)
&=& (\underline{p}_\alpha+\underline{p}^\prime_\alpha)^\nu \,
f(Q^2,s)\nonumber\\&& \hspace{0.5cm}+
(\underline{p}_e+\underline{p}^\prime_e)^\nu \, g(Q^2,s)\, .
\end{eqnarray}
This decomposition is valid in any inertial frame. The problems with
cluster separability do not only modify the covariant structure of
the current, they also affect the form factors associated with the
covariants. As we have indicated in the notation, these form factors
do not only depend on the squared 4-momentum transfer at the
photon-meson vertex
$Q^2=-(\underline{p}_\alpha - \underline{p}^\prime_\alpha)^2$ but also
on Mandelstam $s=(\underline{p}_e+\underline{p}_\alpha)^2$, i.e. the
square of the invariant mass of the electron-meson system.

The necessity of non-physical covariants and corresponding spurious
form factors in our approach resembles the occurrence of analogous
contributions within the covariant light-front formulation of
Carbonell et al.~\cite{Carbonell:1998rj}. Whereas our unphysical
covariant, the sum of the incoming and outgoing electron 4-momenta
$(\underline{p}_e+\underline{p}^\prime_e)$, is caused by wrong
cluster properties inherent in the Bakamjian-Thomas construction,
their unphysical covariant is proportial to a 4-vector $\omega$.
$\omega$ specifies the orientation of the light front and has to be
introduced to render the front-form approach manifestly covariant.

The size of cluster-separability-violating effects can be studied
numerically. To this end (and also for later purposes) we take a
simple harmonic-oscillator wave function
\begin{equation}\label{eq:wavefunc}
 \psi(\kappa)=
 \frac{2}{\pi^{\frac{1}{4}}a^{\frac{3}{2}}}
 \exp\left(-\frac{\kappa^2}{2a^2}\right)\,.
\end{equation}
For further comparison we have chosen the oscillator parameter as
well as the constituent-quark masses to be the same as in
Ref.~\cite{Cheng:1996if}, where form factors of heavy light-mesons
were calculated within the front-form approach. For all heavy-light
mesons, which we will consider in the following, the oscillator
parameter is $a=0.55$~GeV.  The constituent-quark masses are
$m_u=m_d=0.25$~GeV, $m_c=1.6$~GeV and $m_b=4.8$~GeV,
respectively. Since our form factors are only functions of Lorentz
invariants they can be extracted in any inertial frame.\footnote{There is one exception, namely the Breit frame. This frame
corresponds to backward scattering in the electron-meson
CM system. In this frame the two covariants become
proportional and the form factors cannot be uniquely separated}
We choose a center-of-momentum frame in which $\vec{v}=\vec{0}$,
i.e.
$\underline{\vec{p}}^{(\prime)}_\alpha=\underline{\vec{k}}^{(\prime)}_\alpha$,
with
\begin{equation}\label{eq:momentumscatt}
\underline{\vec{k}}_\alpha=-\underline{\vec{k}}_e=
\left( \begin{array}{c} -\frac{Q}{2}\\ 0 \\
\sqrt{\kappa_\alpha^2-\frac{Q^2}{4}}
\end{array}\right)\,  \quad \hbox{and}\quad \vec{q}=\left(
\begin{array}{c} -Q\\ 0 \\ 0 \end{array}\right)\, ,
\end{equation}
where $\kappa_\alpha=|\underline{\vec{k}}_\alpha|=|\underline{\vec{k}}_\alpha^\prime|$.
In this parametrization the modulus of the CM momentum is subject to the
constraint that $\kappa_\alpha^2\geq Q^2/4$, which
means that $s\geq
m_\alpha^2+m_e^2+Q^2/2+2\sqrt{m_\alpha^2+Q^2/4}\sqrt{m_e^2+Q^2/4}$.
The only non-vanishing components of $\tilde{J}_{[\alpha]}^\nu$ in
this frame are $\tilde{J}_{[\alpha]}^0$ and $\tilde{J}_{[\alpha]}^3$
from which we can extract the form factors $f(Q^2,s)$ and $g(Q^2,s)$
by means of Eq.~(\ref{eq:Jemcovdec}) inserting our microscopic
expression, Eqs.~(\ref{eq:voptos}) and (\ref{eq:JQ}), for
$\tilde{J}_{[\alpha]}^\nu$ on the left-hand side.
\begin{figure*}[ht!]
\includegraphics[width=0.45\textwidth]{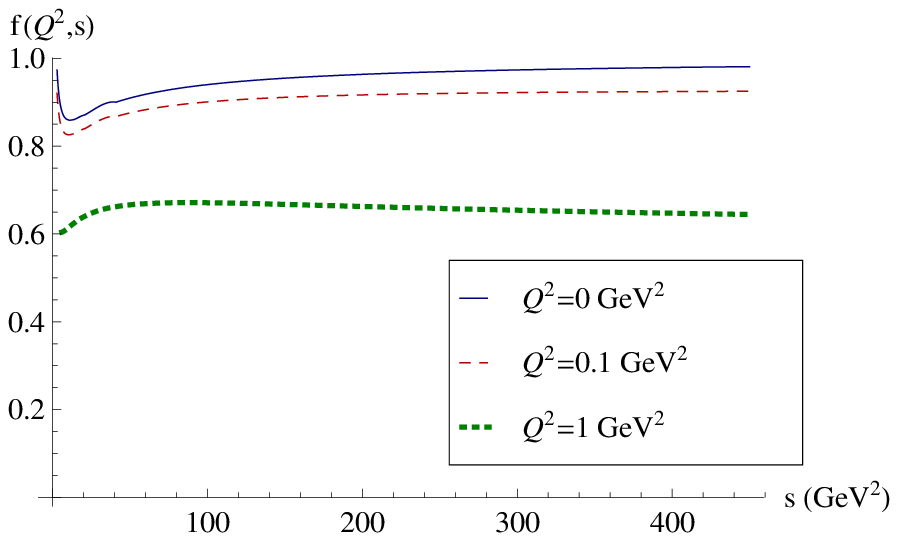}
\includegraphics[width=0.45\textwidth]{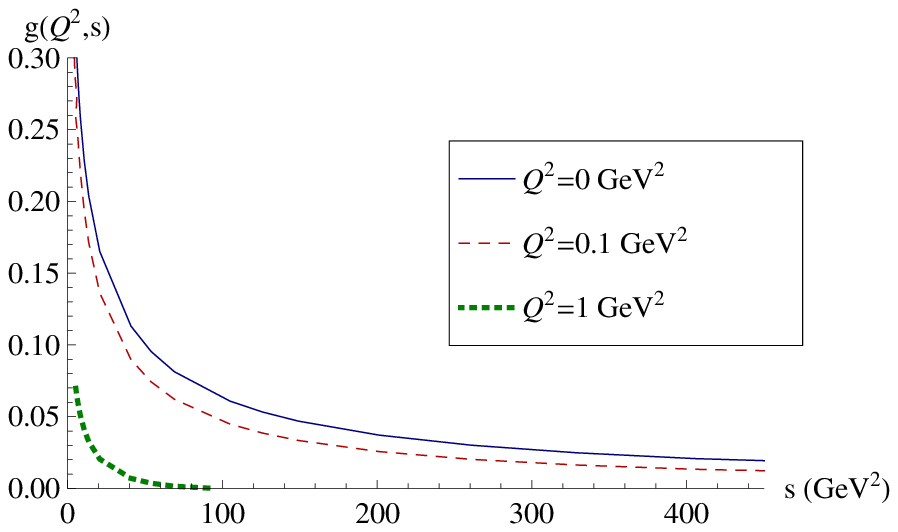}\\
\includegraphics[width=0.45\textwidth]{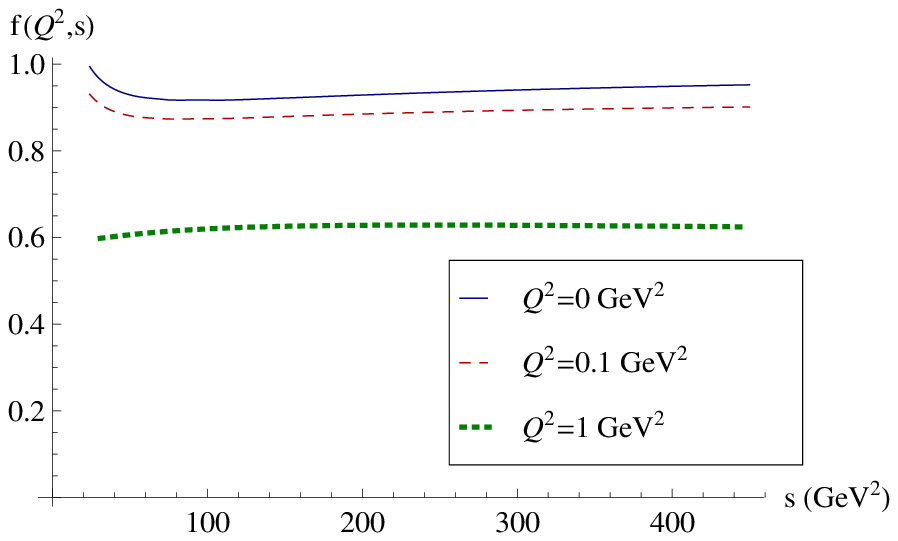}
\includegraphics[width=0.45\textwidth]{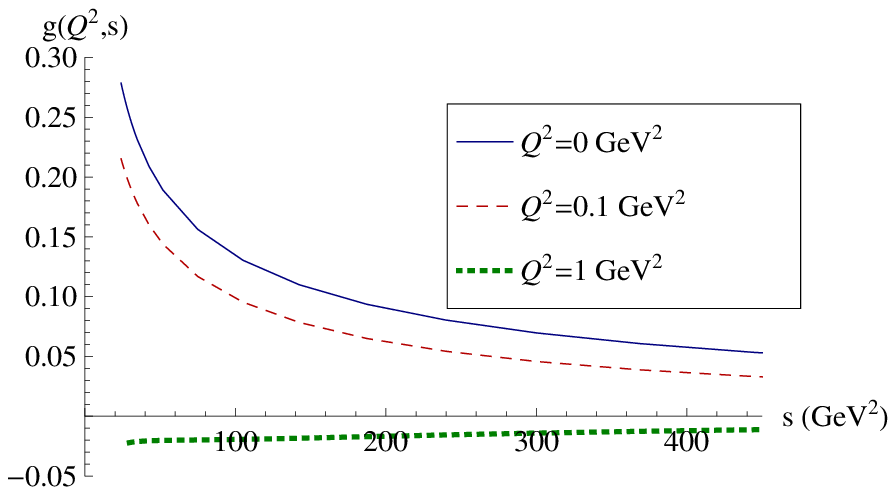}
\caption{(Color online) Mandelstam-$s$ dependence of the physical
and spurious $D^+$ (first row) and $B^-$ (second row)
electromagnetic form factors $f(Q^2,s)$ and $g(Q^2, s)$,
respectively, for different values of $Q^2$ ($0$~GeV$^2$ solid,
$0.1$~GeV$^2$ dashed, $1$~GeV$^2$ dotted) calculated with the oscillator wave function, Eq.~(\ref{eq:wavefunc}), and (mass) parameters given in the sequel.}\label{fig:sdep}
\end{figure*}

The Mandelstam-$s$ dependence of these form factors for a few values
of the momentum transfer $Q^2$ is plotted in Fig.~\ref{fig:sdep} for
$D^+$ and $B^-$ mesons, respectively. What we observe is that the
spurious form factor $g(Q^2,s)$ goes to zero for $s\rightarrow
\infty$ and that the $s$-dependence of the physical form factor
$f(Q^2,s)$ vanishes with increasing $s$. It is therefore suggestive
to take the $s\rightarrow \infty$ limit to get rid of
cluster-separability violating effects and obtain sensible results
for the physical form factors. Taking the $s\rightarrow \infty$
limit means that one extracts the form factor in the infinite
momentum frame of the meson. Not surprisingly, for light-light
systems the resulting analytical expression for the electromagnetic
form factor of a pseudoscalar meson is then seen to be equivalent
with the usual front-form result, obtained from a one-body current
in the $q^+=0$ frame~\cite{Biernat:2009my}. For heavy-light systems
the situation becomes more intricate. Looking more closely at the
form factors for $D^+$ and $B^-$ (cf. Fig.~\ref{fig:sdep}) we
observe that the rate of convergence to the $s\rightarrow \infty$
limit decreases with increasing heavy-quark mass. In order to
extract sensible results for the Isgur-Wise function one thus has to
be very careful when taking the heavy-quark limit $m_Q\rightarrow
\infty$.

\subsection{\label{subsec:decayff}Decay form factors}

\subsubsection{P $\rightarrow$ P transition}
As in the electromagnetic case a weak pseudoscalar-to-pseudoscalar
transition current with the correct transformation properties under
Lorentz transformations is obtained from Eq.~(\ref{eq:Jwkpsps}) by
applying the canonical boost $B_c(\underline{v})$ that connects
physical momenta with CM momenta:
\begin{equation}\label{eq:Jphyspsps}
J^\nu_{B\rightarrow
D}(\vec{\underline{p}}_D^\prime;\vec{\underline{p}}_B):=
[B_c(\underline{v})]^\nu_{\phantom{\nu}\rho} J_{B\rightarrow
D}^\rho(\vec{\underline{k}}_D^\prime;\vec{\underline{k}}_B)\, .
\end{equation}
An appropriate covariant decomposition of this 4-vector current
which holds for arbitrary values of $\vec{\underline{p}}_B$ and
$\vec{\underline{p}}_D^\prime$ takes on the
form~\cite{Wirbel:1985ji}
\begin{eqnarray}\label{eq:Jpspsphys}
\lefteqn{J^\nu_{B\rightarrow
D}(\vec{\underline{p}}_D^\prime;\vec{\underline{p}}_B)}\nonumber\\
&& =\left( (\underline{p}_B+\underline{p}_D')^\nu -
\frac{m^2_B-m^2_D}{\underline{q}^2 }\, \underline{q}^\nu
\right)F_1(\underline{q}^2)\nonumber\\& & \phantom{=} +
\frac{m_B^2-m_D^2}{\underline{q}^2}\, \underline{q}^\nu
F_0(\underline{q}^2)\, ,
\end{eqnarray}
with the time-like 4-momentum transfer
$\underline{q}=(\underline{p}_B-\underline{p}_D)$. Unlike the
electromagnetic case wrong cluster properties of the
Bakamjian-Thomas construction do not entail unphysical properties of
the weak decay current $J^\nu_{B\rightarrow
D}(\vec{\underline{p}}_D^\prime;\vec{ \underline{p}}_B)$ if the
microscopic expression, Eq.~(\ref{eq:Jwkpsps}), is inserted on the
right-hand-side of Eq.~(\ref{eq:Jphyspsps}). One neither needs
additional unphysical covariants to span the 4-vector
$J^\nu_{B\rightarrow D}$, nor do the form factors exhibit a
dependence on Lorentz invariants different
$\underline{q}^2$.\footnote{One could think of two additional
unphysical covariants (a vector and an axial-vector) constructed
with the 4-vector $(\underline{k}_e - \underline{k}_{\bar{\nu}_e})$
and an additional dependence of the form factors on
$(\underline{k}_e+\underline{k}_D)^2$.}

The finding that wrong cluster properties of the Bakamjian-Thomas
construction do not have obvious physical consequences for the weak
decay current $J^\nu_{B\rightarrow D}$, whereas they lead to
unphysical features of the electromagnetic current
$\tilde{J}_{[\alpha]}^\nu$, has essentially three reasons:
\begin{itemize}
\item[i)]Only the final state of the decay process is affected by wrong
cluster properties, since the initial state is just the confined
quark-antiquark pair with no additional particle present. In
electron scattering off a bound system the presence of the electron
modifies the bound-state wave function in both, the initial and the
final states.

\item[ii)]There is no constraint from current conservation for
the decay current $J^\nu_{B\rightarrow D}$ such that both 4-vectors,
$(\underline{p}_B+\underline{p}_D')$ and
$\underline{q}=(\underline{p}_B-\underline{p}_D')$, can be used to
express the decay current (cf. Eq.~(\ref{eq:Jpspsphys})). As it
turns out, this suffices. The electromagnetic current
$\tilde{J}_{[\alpha]}^\nu$, on the other hand, is conserved. It thus
cannot have a component into the direction of the momentum transfer
$(\underline{p}_\alpha-\underline{p}_\alpha')$, but is also not just
proportional to $(\underline{p}_\alpha+\underline{p}_\alpha')$
alone. Therefore one is forced to introduce the unphysical covariant
$(\underline{p}_e+\underline{p}_e')$.

\item[iii)] Both, the electromagnetic and the weak form factors are
functions of $|\vec{\underline{q}}|$, the modulus of the 3-momentum
transfer between the meson in the incoming and outgoing state. Since
form factors are frame independent quantities it should be possible
to express $|\vec{q}|$ in terms of Lorentz-invariant quantities. In the
case of the weak decay $|\vec{\underline{q}}|^2$ and
$\underline{q}_\mu \underline{q}^\mu$ are directly related (see
below). In the case of electron scattering one needs in general
Mandelstam $s$ and $t=\underline{q}_\mu \underline{q}^\mu$ to
express $|\vec{q}|^2$. This is the reason why the weak form factors
can be written as functions of $\underline{q}_\mu \underline{q}^\mu$
only, whereas the electromagnetic form factors exhibit an additional
(unphysical) dependence on Mandelstam $s$.

\end{itemize}
The observation that $J^\nu_{B\rightarrow D}$ does not exhibit
unphysical features does not necessarily mean that there are no
problems with wrong cluster properties within our approach in the
decay process. As mentioned above wrong cluster properties could
still affect the wave function of the final state. But, unlike the electromagnetic case, there is no
simple way to separate corresponding contributions in the decay
current.\footnote{Formally, cluster separability can be restored by
means of packing operators~\cite{Keister:1991sb}. Practically such
packing operators are hard to construct, in particular for a
multichannel mass operator.} The emergence of heavy-quark symmetry, which relates electromagnetic and weak decay form factors, however, will let us conclude that such wrong cluster properties become negligible in the heavy-quark limit.

Equation~(\ref{eq:Jpspsphys}) is a general representation for the
weak decay current which holds in any inertial frame. A convenient
choice for the extraction of the decay form factors
$F_0(\underline{q}^2)$ and $F_1(\underline{q}^2)$ is the CM frame
($\vec{v}=\vec{0}$) in which
\begin{equation}\label{eq:decaykinem}
\underline{k}_B= \left( \begin{array}{c} m_B\\ 0 \\0\\0
\end{array}\right) \quad\hbox{and}\quad
\underline{k}_D'= \left( \begin{array}{c} \sqrt{m_D^2+\kappa_D^{2}}\\
\kappa_D
\\0\\0
\end{array}\right)\end{equation}
with
\begin{equation}
\kappa_D^2=\frac{1}{4m_B^2}
(m_B^2+m_D^2-\underline{q}^2)^2-m_D^2\,.
\end{equation}

The modulus of the $D$ meson CM momentum $\kappa_D=|\underline{\vec{k}}_D^\prime|$ is thus
restricted by $0\leq \kappa_D^2 \leq
(m_B^2-m_D^2)^2/(4m_B^2)$. As in the electromagnetic case the momentum is transferred in $x$-direction. The allowed values of the 4-momentum
transfer squared are
\begin{equation}
0\leq \underline{q}^2 \leq (m_B-m_D)^2 \, .
\end{equation}
The $\nu=2,3$ components of the weak transition current
$J^\nu_{B\rightarrow D}$ vanish in this kinematics. As it should be, the non-zero $\nu=0,1$ components of $J^\nu_{B\rightarrow D}$ are solely determined by the vector part ($\propto \gamma^\nu$) of the $Wbc$-vertex. The axial-vector part ($\propto \gamma^\nu \gamma^5$) of the vertex does not contribute to the $B\rightarrow D$ transition. The form factors
$F_0(\underline{q}^2)$ and $F_1(\underline{q}^2)$ can be determined
uniquely by projecting onto the corresponding 4-vectors:
\begin{equation}\label{eq:F0}
\hspace{-2cm} F_0(\underline{q}^2)=\frac{1}{m_B^2-m_D^2}\, \underline{q}_\nu
J_{B\rightarrow
D}^\nu(\vec{\underline{k}}_D^\prime;\vec{\underline{k}}_B)\, ,
\end{equation}
\begin{eqnarray}\label{eq:F1}\lefteqn{
F_1(\underline{q}^2)=-\frac{\underline{q}^2}{4 m_B^2
m_D^2}\left[\left(\frac{m_B^2+m_{D}^2-\underline{q}^2}{2 m_B m_{D}}
\right)^2-1\right]^{-1}}\\&&\times \left(
(\underline{p}_B+\underline{p}_D')_\nu -
\frac{m^2_B-m^2_D}{\underline{q}^2 }\, \underline{q}_\nu \right)
J_{B\rightarrow
D}^\nu(\vec{\underline{k}}_D^\prime;\vec{\underline{k}}_B)\,
.\nonumber
\end{eqnarray}
The constraint $F_0(0)=F_1(0)$, that eliminates the spurious pole at $\underline{q}^2=0$, is automatically satisfied for the form factors calculated from our transition current, Eq.~(\ref{eq:Jwkpsps}).

\subsubsection{P $\rightarrow$ V transition}
The weak pseudoscalar-to-vector transition current with the correct
transformation properties under Lorentz transformations is obtained
from Eq.~(\ref{eq:Jwkpsv}) by applying again the canonical boost
$B_c(\underline{v})$ that connects physical momenta with CM momenta.
Linked with this boost is a Wigner rotation of the vector-meson
spin:
\begin{eqnarray}\label{eq:Jphyspsv}
\lefteqn{J^\nu_{B\rightarrow
D^\ast}(\vec{\underline{p}}_{D^\ast}^{\prime},\underline\sigma^\prime_{D^\ast};
\vec{\underline{p}}_B):=
[B_c(\underline{v})]^\nu_{\phantom{\nu}\rho} J_{B\rightarrow
D^\ast}^\rho(\vec{\underline{k}}_{D^\ast}^{\prime},\underline{\mu}^\prime_{D^\ast};
\vec{\underline{k}}_B)}\nonumber\\&& \hspace{2.5cm}\times
D^{1\ast}_{\underline{\mu}^\prime_{D^\ast}
\underline\sigma^\prime_{D^\ast}}\left[R^{-1}_{\mathrm{W}}\!
\left({\underline{k}}_{D^\ast}^{\prime}/m_{D^\ast},
B_c(v)\right)\right]\, . \nonumber\\
\end{eqnarray}
A common covariant decomposition of this 4-vector current has the form~\cite{Wirbel:1985ji}
\begin{eqnarray}
\lefteqn{J^\nu_{B\rightarrow
D^\ast}(\vec{\underline{p}}_{D^\ast}^{\prime},\underline\sigma^\prime_{D^\ast};
\vec{\underline{p}}_B)}\nonumber\\&=& \frac{2 i\epsilon^{\nu\mu\rho\sigma}}{m_B+m_{D^*}}\,\epsilon^*_\mu(\vec{\underline{p}}_{D^\ast}^{\prime}, \underline\sigma^\prime_{D^\ast})\, \underline{p}'_{D^\ast\rho}\, \underline{p}_{B\sigma} \, V(\underline{q}^2)\nonumber\\ && - (m_B+m_{D^*})\, \epsilon^{*\nu}(\vec{\underline{p}}_{D^\ast}^{\prime}, \underline\sigma^\prime_{D^\ast})\, A_1(\underline{q}^2)\nonumber\\&& + \frac{\epsilon^*(\vec{\underline{p}}_{D^\ast}^{\prime}, \underline\sigma^\prime_{D^\ast}) \cdot \underline{q}}{m_B+m_{D^*}}\,(\underline{p}_B+\underline{p}_{D^\ast}')^\nu\, A_2(\underline{q}^2)\nonumber\\&& +2 m_{D^*}\,\frac{\epsilon^*(\vec{\underline{p}}_{D^\ast}^{\prime}, \underline\sigma^\prime_{D^\ast}) \cdot \underline{q}}{\underline{q}^2}\, \underline{q}^\nu \, A_3(\underline{q}^2)\nonumber\\
&&- 2m_{D^*}\,\frac{\epsilon^*(\vec{\underline{p}}_{D^\ast}^{\prime}, \underline\sigma^\prime_{D^\ast}) \cdot \underline{q}}{\underline{q}^2}\, \underline{q}^\nu\, A_0(\underline{q}^2)\, ,
\end{eqnarray}
with $\epsilon^*(\vec{\underline{p}}_{D^\ast}^{\prime}, \underline\sigma^\prime_{D^\ast})$ being the polarization 4-vector of the $D^\ast$ meson and $A_3(\underline{q}^2)$ the linear combination
\begin{equation}
A_3(\underline{q}^2)= \frac{m_B+m_{D^\ast}}{2 m_{D^\ast}}\, A_1(\underline{q}^2) - \frac{m_B-m_{D^\ast}}{2 m_{D^\ast}} A_2(\underline{q}^2)\, .
\end{equation}
The constraint $A_3(0)=A_0(0)$, that holds automatically for the form factors calculated from our transition current, Eq.~(\ref{eq:Jwkpsv}), guarantees that there is no pole at $\underline{q}^2=0$. As for the $B\rightarrow D$ transition wrong cluster properties of the Bakamjian-Thomas construction do not lead to unphysical features of the $J^\nu_{B\rightarrow D^\ast}$ decay current. The vector $V(\underline{q}^2)$ and the axial-vector form factors $A_i(\underline{q}^2)$ are determined by the vector part ($\propto \gamma^\nu$) and the axial-vector part ($\propto \gamma^\nu \gamma^5$) of the $Wbc$-vertex, respectively.

Taking the same kinematics as for the $B\rightarrow D$ decay (cf. Eq.~(\ref{eq:decaykinem})) the polarization vectors $\epsilon(\underline{\vec{k}}^\prime_{D^\ast},\underline{\mu}_{D^\ast}^\prime)$ are given by:
\begin{eqnarray}\label{eq:Dpol}
\epsilon(\underline{\vec{k}}^\prime_{D^\ast},\pm 1)&=&\frac{1}{\sqrt{2}}(\mp \frac{\kappa_{D^\ast}}{m_{D^\ast}}, \mp\sqrt{1+(\frac{\kappa_{D^\ast}}{m_{D^\ast}})^2},-i,0)\, , \nonumber\\ \epsilon(\underline{\vec{k}}^\prime_{D^\ast},0 )&=&(0,0,0,1)\, .
\end{eqnarray}
This kinematics leads to 10 non-vanishing current matrix elements
$J^2(0)$, $J^3(0)$, $J^\mu(\pm 1)$, $\mu=0,1,2,3$.  Here we have
introduced the short-hand notation
$J^\nu(\underline\mu^\prime_{D^\ast}) := J^\nu_{B\rightarrow
D^\ast}(\vec{\underline{k}}_{D^\ast}^\prime,
\underline\mu^\prime_{D^\ast};\vec{\underline{k}}_B)$. $J^\mu(1)$
and $J^\mu(-1)$ are related by space reflection. We are thus left
with 6 current matrix elements with only 4 of them being
independent. The form factors $A_0$ and $A_2$ enter only $J^0(1)$
and $J^1(1)$. $J^2(0)$, $J^3(0)$, $J^0(1)$ and $J^1(1)$ constitute
thus an appropriate set of current matrix elements from which all
the $P\rightarrow V$ decay form factors can be extracted. Instead of
solving the linear equations which relate the form factors to the
current matrix elements $J^\nu(\underline{\mu}^\prime_{D^\ast})$ we express the
form factors again in terms of appropriate projections:
\begin{eqnarray}\label{eq:V}
V(\underline{q}^2)&=&\frac{i(m_B+m_{D^\ast})}{2 m_B^2 m_{D^\ast}^2}\left[\left(\frac{m_B^2+m_{D^\ast}^2-
\underline{q}^2}{2 m_B m_{D^\ast}} \right)^2-1\right]^{-1}\nonumber\\
&&\times \epsilon_\mu(\vec{\underline{k}}_{D^\ast}^{\prime}, \underline{\mu}^\prime_{D^\ast}=0)\,
\underline{k}'_{D^\ast\rho}\, \underline{k}_{B\sigma} \, \nonumber\\&&\times
\epsilon_{\nu}^{\phantom{\nu}\mu\rho\sigma} J^\nu_{B\rightarrow
D^\ast}(\vec{\underline{k}}_{D^\ast}^{\prime},\underline{\mu}^\prime_{D^\ast}=0; \vec{\underline{k}}_B)\, ,\\
A_0(\underline{q}^2)&=&\frac{1}{\sqrt{2} m_B m_{D^\ast}
}\left[\left(\frac{m_B^2+m_{D^\ast}^2- \underline{q}^2}{2 m_B
m_{D^\ast}} \right)^2-1\right]^{-1/2}\nonumber\\&&\times
\underline{q}_\nu J^\nu_{B\rightarrow
D^\ast}(\vec{\underline{k}}_{D^\ast}^{\prime},\underline{\mu}^\prime_{D^\ast}=1;
\vec{\underline{k}}_B)\, ,\\
A_1(\underline{q}^2)&=&\frac{1}{m_B+m_{D^\ast}}\,
\epsilon_\nu(\vec{\underline{k}}_{D^\ast}^{\prime},
\underline{\mu}^\prime_{D^\ast}=0)\, \nonumber\\ & &\times J^\nu_{B\rightarrow
D^\ast}(\vec{\underline{k}}_{D^\ast}^{\prime},\underline{\mu}^\prime_{D^\ast}=0;
\vec{\underline{k}}_B)\, .
\end{eqnarray}
The expression for $A_2(\underline{q}^2)$ is a little bit more
complicated:
\begin{eqnarray}\label{eq:A2}
A_2(\underline{q}^2)\!&=&\!\!\frac{\underline{q}^2
(m_B+m_{D^\ast})}{4 m_B^2 m_{D^\ast}^2
}\left[\left(\frac{m_B^2+m_{D^\ast}^2- \underline{q}^2}{2 m_B
m_{D^\ast}} \right)^2\!\!\!\!-1\right]^{-1}\nonumber\\
&&\hspace{-1cm}\times\Bigg\{
\frac{\sqrt{2}}{m_B}\left[\left(\frac{m_B^2+m_{D^\ast}^2-
\underline{q}^2}{2 m_B m_{D^\ast}}
\right)^2\!\!\!\!-1\right]^{-1/2}\nonumber\\&&\hspace{-1cm}
\phantom{\times\Bigg\{}\times\left(
(\underline{p}_B+\underline{p}_{D^\ast}') -
\frac{m^2_B-m^2_{D^\ast}}{\underline{q}^2 }\, \underline{q}
\right)_\nu\nonumber\\&&\hspace{-1cm}\phantom{\times\Bigg\{}\times
J_{B\rightarrow
D^\ast}^\nu(\vec{\underline{k}}_{D^\ast}^{\prime},\underline{\mu}^\prime_{D^\ast}=1;
\vec{\underline{k}}_B)\nonumber\\
&&\hspace{-1cm}\phantom{\times} - \left[ 1- \frac{m_B^2 -
m_{D^\ast}^2}{\underline{q}^2}
\right]\epsilon_\nu(\vec{\underline{k}}_{D^\ast}^{\prime},
\underline{\mu}^\prime_{D^\ast}=0 )\nonumber\\
&& \hspace{-1cm}\phantom{\times\Bigg\{} \times
 J^\nu_{B\rightarrow
D^\ast}(\vec{\underline{k}}_{D^\ast}^{\prime},\underline{\mu}^\prime_{D^\ast}=0;
\vec{\underline{k}}_B) \Bigg\}\, .\nonumber\\\end{eqnarray}

Having derived analytical expressions for the electromagnetic and
weak currents and form factors  we are now going to study their
properties in the heavy-quark limit.

\section{\label{sec:heavyqlim}The heavy-quark limit}
In the heavy-quark limit the masses of the heavy quarks and,
consequently, the masses  of the heavy hadrons are sent to infinity.
This leads to additional symmetries which will be discussed later.
With the hadron masses also their momenta go to infinity. What,
however, stays finite is the product $\underline{v}_\alpha\cdot
\underline{v}_{\alpha^{(\prime)}}^{\prime}$ of the hadron 4-velocities. One is then interested in  the dependence of form factors on the (finite) velocity product $\underline{v}_\alpha\cdot
\underline{v}_{\alpha^{(\prime)}}^{\prime}$. Thus it
makes more sense to characterize the state of a heavy hadron by its
velocity rather than by its momentum. To be more precise, the limit
$m_Q\rightarrow \infty$ has to be taken in such a way that
\begin{equation}
\underline{v}_\alpha\cdot \underline{v}_{\alpha^{(\prime)}}
^{\prime}=\frac{\underline{k}_\alpha\cdot
\underline{k}_{\alpha^{(\prime)}}^\prime}{m_\alpha
m_{\alpha^{(\prime)}}}
\end{equation}
stays constant. In this limit both, the binding energy and the light-quark mass, become negligible, i.e.
\begin{equation}
m_{Q^{(\prime)}} = m_{\alpha^{(\prime)}} \quad \hbox{and} \quad \frac{m_q}{m_{Q^{(\prime)}}}=0 \quad \hbox{for} \quad m_{Q^{(\prime)}}\rightarrow \infty\, .
\end{equation}
Furthermore it is assumed that the meson wave functions do not depend on the flavor of the heavy quarks when the masses of the heavy quarks go to infinity. This is our precise definition of the \lq\lq heavy-quark limit\rq\rq\, (h.q.l.).

\subsection{Space-like momentum transfer}\label{subsec:IWspacel}
Let us start with the heavy-quark limit of the electromagnetic
pseudoscalar-meson current
$\tilde{J}_{[\alpha]}^\nu(\vec{\underline{k}}_\alpha^{\,\prime}
;\vec{\underline{k}}_\alpha)$ (cf. Eqs.~(\ref{eq:voptos}) and
(\ref{eq:JQ})). The first step towards the heavy-quark limit is to
express the meson momenta and the momenta of the heavy quarks in
terms of velocities. To this aim we note that
\begin{equation}\label{eq:Qemvv}
Q=\sqrt{-(\underline{k}_\alpha-\underline{k}_\alpha^\prime)^2}=
2m_\alpha\sqrt{\frac{\underline{v}_\alpha\cdot \underline{v}_\alpha^{\prime}-1}{2}}=:
2m_\alpha u\, .
\end{equation}
This means, in particular, that not only the heavy-quark mass, but
also the momentum transfer goes to infinity, when taking the
heavy-quark limit. As a consequence
$\tilde{J}_{\bar{q}}^\nu(\vec{\underline{k}}_\alpha^{\,\prime}
;\vec{\underline{k}}_\alpha)$, the part of the current in which the
momentum is transferred to the light antiquark, vanishes. The formal
reason is that the wave-function overlap vanishes (exponentially)
when the light antiquark has to absorb an infinite amount of
momentum. It thus remains to investigate the heavy-quark limit of
$\tilde{J}_{Q}^\nu(\vec{\underline{k}}_\alpha^{\,\prime}
;\vec{\underline{k}}_\alpha)$, i.e. the part of the current in which
the momentum is  transferred to the heavy quark. Taking the
parametrization of meson momenta that has been defined in
Eq.~(\ref{eq:momentumscatt}) and going over to velocities we have:
\begin{eqnarray}\label{eq:vscatt}
\underline{k}_\alpha=m_\alpha {\begin{pmatrix}
\sqrt{1+\nu_\alpha^2}\\-u\\0\\\sqrt{\nu_\alpha^2-u^2}\end{pmatrix}}=m_\alpha
{\underline{v}_\alpha}\,,
\nonumber\\\phantom{a}\\ \underline{k}_\alpha^\prime=m_\alpha
{\begin{pmatrix}
\sqrt{1+\nu_\alpha^2}\\u\\0\\\sqrt{\nu_\alpha^2-u^2}\end{pmatrix}}=m_\alpha
{\underline{v}_\alpha^\prime} \, ,\nonumber
\end{eqnarray}
where
$\nu_\alpha=|\underline{\vec{v}}_\alpha|=|\underline{\vec{v}}_\alpha^\prime|$
and $u$ is the  shorthand notation introduced in
Eq.~(\ref{eq:Qemvv}). The modulus of the 4-momentum transfer squared
$Q^2=-q_\mu q^\mu$ and our new variable $\underline{v}_\alpha\cdot
\underline{v}_\alpha^{\prime}$ are then related by
\begin{equation}\label{eq:q2scatt}
Q^2=2 m_\alpha^2(\underline{v}_\alpha\cdot
\underline{v}_\alpha^{\prime}-1)
\end{equation}
which means that
$$
\underline{v}_\alpha\cdot \underline{v}_\alpha^{\prime}\geq 1
$$
for elastic electron-meson scattering. Using now that
\begin{widetext}
\begin{eqnarray}\label{eq:hqlmomenta}
&&\underline{\vec{k}}_\alpha^{\,(\prime)}\, ,
\vec{k}_Q^{\,(\prime)}\, \stackrel{\mathrm{h.q.l.}}{\longrightarrow}
\, m_\alpha\, \underline{\vec{v}}_\alpha^{\,(\prime)}\,
,\quad\frac{|\vec{k}_{\bar q}^{\, (\prime)}|}{m_Q}\,
,\frac{|\vec{\tilde{k}}_{\bar q}^{\, (\prime)}|}{m_Q}\, ,
\frac{|\vec{\tilde{k}}_{Q}^{\,
(\prime)}|}{m_Q}\stackrel{\mathrm{h.q.l.}}{\longrightarrow} 0\, ,
\quad\hbox{and}\quad {\vec{v}}_{Q\bar{q}}^{\,(\prime)}\stackrel{\mathrm{h.q.l.}}{\longrightarrow} \underline{\vec{v}}_{\alpha}^{\,(\prime)}\, ,
\end{eqnarray}
the Wigner rotations of the heavy-quark spin become the identity in the heavy-quark limit and the kinematical factors in the pseudoscalar meson current $\tilde{J}_{Q}^\nu(\vec{\underline{k}}_\alpha^{\,\prime}
;\vec{\underline{k}}_\alpha)$ (cf. Eq.~(\ref{eq:JQ})) simplify considerably:
\begin{eqnarray}\label{eq:hqljem}
J^{\nu}_Q
(\vec{\underline{k}}_\alpha^\prime,\vec{\underline{k}}_\alpha)
\quad\stackrel{\mathrm{h.q.l.}}{\longrightarrow}\quad m_\alpha
\tilde{J}^{\nu}_\infty
(\vec{\underline{v}}_\alpha^\prime,\vec{\underline{v}}_\alpha)&=&
m_\alpha \int\, \frac{d^3\tilde{k}_{\bar{q}}^\prime}{4\pi}\,
\sqrt{\frac{\omega_{\tilde{k}_{\bar{q}}}}
{\omega_{\tilde{k}^\prime_{\bar{q}}}}}
 \,  \bigg\{\!\sum_{\mu_Q,\mu_Q^\prime
=\pm \frac{1}{2}}\!\!\!
\bar{q}_{\mu_Q^\prime}(\vec{\underline{v}}_\alpha^{\,\prime})\,\gamma^\nu\,
u_{\mu_Q}(\vec{\underline{v}}_\alpha) \\ && \times\frac{1}{2}\,
D^{1/2}_{\mu_Q\mu_Q^\prime}\!
\left[\!R^{-1}_{\mathrm{W}}\!\left(\frac{\tilde{k}_{\bar{q}}}{m_{\bar{q}}},
B_c(\underline{v}_{\alpha})\right)\,
\!R_{\mathrm{W}}\!\left(\frac{\tilde{k}_{\bar q}^\prime}{m_{\bar
q}}, B_c(\underline{v}_{\alpha}^\prime)\right)\, \right] \bigg\}\,
\psi^\ast\,(\vert \vec{\tilde{k}}_{\bar{q}}^\prime\vert)\,  \psi
 \,(\vert \vec{\tilde{k}}_{\bar{q}}\vert)\,\, .\nonumber
\end{eqnarray}
\end{widetext}
One can see immediately that the integrand is independent of the
heavy quark mass.  The only dependencies showing up are those on the
integration variables $\vec{\tilde{k}}_{\bar{q}}^{\,\prime}$~  and
on the meson velocities $\vec{\underline{v}}_{\alpha}^{\,(\prime)}$.
The term within the curly brackets comes from the spin of the
quarks. For spinless quarks it would coincide with the pointlike
current of the pseudoscalar meson
$(\underline{v}_\alpha+\underline{v}_\alpha^\prime)^\nu$. Dropping
this factor the integral on the right-hand side of
Eq.~(\ref{eq:hqljem}) would already give the Isgur-Wise function for
a scalar meson composed of spinless quarks. The general covariant
structure of $\tilde{J}^{\nu}_\infty
(\vec{\underline{v}}_\alpha^\prime,\vec{\underline{v}}_\alpha)$ for
spin-1/2 quarks follows from Eq.~(\ref{eq:Jemcovdec}) by expressing
the momenta in terms of velocities
\begin{eqnarray}\label{eq:jemhqcov}
\tilde{J}^{\nu}_\infty
(\vec{\underline{v}}_\alpha^\prime,\vec{\underline{v}}_\alpha) &=&
(\underline{v}_\alpha+\underline{v}_\alpha^{\,\prime})^\nu
\,\tilde{f}(\underline{v}_\alpha\cdot
\underline{v}_{\alpha}^{\prime},\nu_{\alpha})\nonumber\\&&+
\frac{m_e}{m_\alpha}(\underline{v}_e+\underline{v}_e^{\,\prime})^\nu
\,\tilde{g}(\underline{v}_\alpha\cdot
\underline{v}_{\alpha}^{\prime},\nu_{\alpha})\, ,
\end{eqnarray}
where
\begin{equation}
\frac{m_e}{m_\alpha}(\underline{v}_e+\underline{v}_e^{\,\prime})=2
(\nu_\alpha,0,0,\sqrt{\nu_\alpha^2-u^2})\, .
\end{equation}
\begin{figure*}[ht!]
\includegraphics[width=0.45\textwidth]{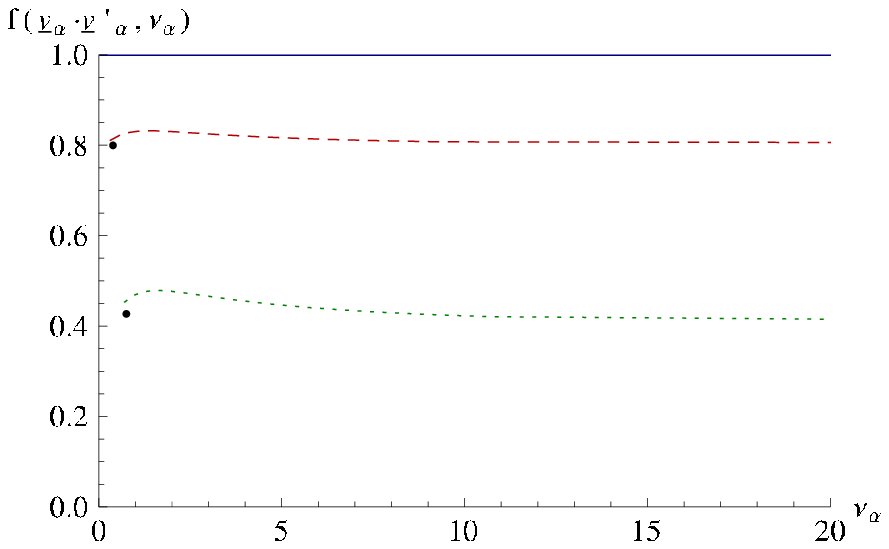}\hspace{1cm}
\includegraphics[width=0.45\textwidth]{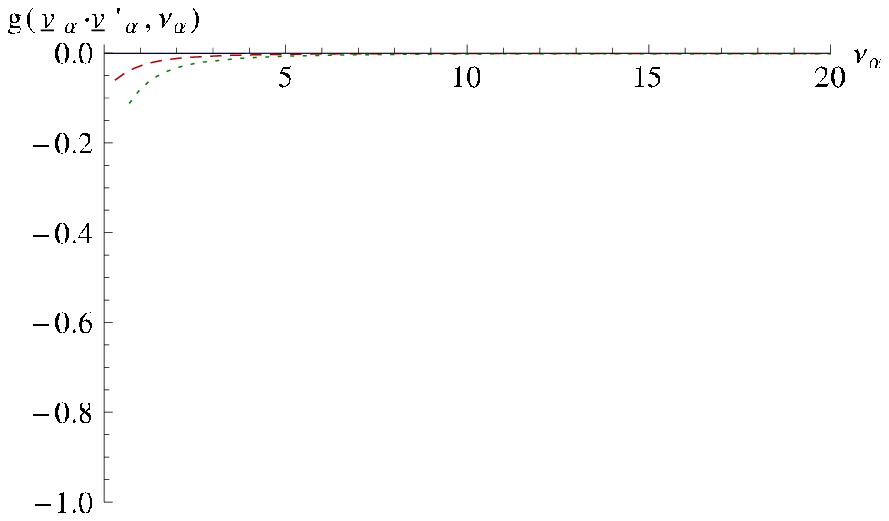}
\caption{(Color online) Physical and spurious electromagnetic form
factors, $\tilde{f}(\underline{v}_\alpha\cdot
\underline{v}_{\alpha}^{\prime},\nu_{\alpha})$ and
$\tilde{g}(\underline{v}_\alpha\cdot
\underline{v}_{\alpha}^{\prime},\nu_{\alpha})$, of a heavy-light
pseudoscalar meson in the heavy-quark limit with the model parameters being the same as in Fig.~\ref{fig:sdep}. Their dependence on the
modulus of the meson velocities $\nu_\alpha$ is plotted for
different values of $\underline{v}_\alpha\cdot
\underline{v}_{\alpha}^{\prime}$ ($1$ solid, $1.2$ dashed, $2$
dotted). The black dots in the left figure are the values for the
Isgur-Wise function directly calculated in the Breit frame
($\nu_\alpha=u$), where $\tilde{f}(\underline{v}_\alpha\cdot
\underline{v}_{\alpha}^{\prime},\nu_{\alpha})$ and
$\tilde{g}(\underline{v}_\alpha\cdot
\underline{v}_{\alpha}^{\prime},\nu_{\alpha})$ cannot be separated.}
\label{fig:vdep}
\end{figure*}
As it turns out and as it is indicated in Eq.~(\ref{eq:jemhqcov})
$\tilde{J}^{\nu}_\infty
(\vec{\underline{v}}_\alpha^\prime,\vec{\underline{v}}_\alpha)$
still does not have all the desired properties. Effects of wrong
cluster properties, that are inherent in our approach, do not go
away by taking the heavy-quark limit. It is, in general, not
possible to write $\tilde{J}^{\nu}_\infty
(\vec{\underline{v}}_\alpha^\prime,\vec{\underline{v}}_\alpha)$ as a
product of the covariant
$(\underline{v}_\alpha+\underline{v}_\alpha^{\,\prime})^\nu$ times
the Isgur-Wise function $\xi(\underline{v}_\alpha\cdot
\underline{v}_{\alpha}^{\prime})$. One rather needs a second
covariant built from the electron velocities. In addition, the form
factors are not only functions of $\underline{v}_\alpha\cdot
\underline{v}_{\alpha}^{\prime}$, but exhibit also a dependence on
the modulus of the meson velocities $\nu_\alpha$. The latter
dependence corresponds to the Mandelstam-$s$ dependence
($\nu_\alpha=(\frac{\sqrt{s}}{m_\alpha}-\frac{m_\alpha}{\sqrt{s}})/2$
with
$s=m_\alpha^2(\underline{v}_\alpha+\frac{m_e}{m_\alpha}\underline{v}_e)^2$)
which we have already discussed in Sec.~\ref{sec:currents} (cf.
Fig.~\ref{fig:sdep}) for the case of finite heavy-quark mass and
which also occurs in light-light
systems~\cite{Biernat:2009my,Biernat:2011mp}. The
$\nu_\alpha$-dependence of $\tilde{f}(\underline{v}_\alpha\cdot
\underline{v}_{\alpha}^{\prime},\nu_{\alpha})$ and
$\tilde{g}(\underline{v}_\alpha\cdot
\underline{v}_{\alpha}^{\prime},\nu_{\alpha})$ is displayed in
Fig.~\ref{fig:vdep} for different values of
$\underline{v}_\alpha\cdot \underline{v}_{\alpha}^{\prime}$ with the
wave function of the heavy-light system being the one introduced in
Eq.~(\ref{eq:wavefunc}). One observes that both, the
$\nu_\alpha$-dependence and the spurious form factor
$\tilde{g}(\underline{v}_\alpha\cdot
\underline{v}_{\alpha}^{\prime},\nu_{\alpha})$ vanish rather quickly
with increasing $\nu_\alpha$. It is therefore suggestive to identify
the Isgur-Wise function $\xi(\underline{v}_\alpha\cdot
\underline{v}_{\alpha}^{\prime})$ with the
$\nu_\alpha\rightarrow\infty$ limit of
$\tilde{f}(\underline{v}_\alpha\cdot
\underline{v}_{\alpha}^{\prime},\nu_{\alpha})$. In this limit the
unwanted $\nu_\alpha$-dependence goes away and
$\tilde{J}^{\nu}_\infty(\vec{\underline{v}}_\alpha^\prime,\vec{\underline{v}}_\alpha)$
acquires the expected structure
\begin{equation}
\tilde{J}^{\nu}_\infty(\vec{\underline{v}}_\alpha^\prime,
\vec{\underline{v}}_\alpha)\stackrel{\nu_\alpha\rightarrow\infty}{\longrightarrow}
(\underline{v}_{\alpha}+\underline{v}_{\alpha}^{\prime})^\nu\,
\xi_{\mathrm{IF}}(\underline{v}_\alpha\cdot
\underline{v}_{\alpha}^{\prime})\, ,
\end{equation}
with a simple analytical expression for the Isgur-Wise function
\begin{equation}\label{eq:xiif}
\xi_{\mathrm{IF}}(\underline{v}_\alpha\cdot\underline{v}_{\alpha}^{\prime})=
\int\, \frac{d^3\tilde{k}_{\bar{q}}^\prime}{4\pi}\,
\sqrt{\frac{\omega_{\tilde{k}_{\bar{q}}}}
{\omega_{\tilde{k}^\prime_{\bar{q}}}}}\,\mathcal{S}_{\mathrm{IF}}\,
\psi^\ast\,(\vert \vec{\tilde{k}}_{\bar{q}}^\prime\vert)\,  \psi
 \,(\vert \vec{\tilde{k}}_{\bar{q}}\vert)\,\, .
\end{equation}
Taking the limit $\nu_\alpha\rightarrow\infty$ means that the
$\gamma^\ast M_\alpha\rightarrow M_\alpha$ subprocess is considered
in the infinite-momentum frame of the meson
$M_\alpha$.\footnote{After having performed the heavy-quark limit
the infinite-momentum frame has to be understood as a frame in which
the $3$-components of the incoming and outgoing meson velocities
$\underline{v}_\alpha^{(\prime) 3}$ go to infinity.} This is the
reason why a subscript \lq\lq IF\rq\rq\ is attached to the
Isgur-Wise function and the spin-rotation factor. The relation
between $\omega_{\tilde{k}_{\bar{q}}}$ and
$\omega_{\tilde{k}^\prime_{\bar{q}}}$ (and hence between $\vert
\vec{\tilde{k}}_{\bar{q}}\vert$ and $\vert
\vec{\tilde{k}}_{\bar{q}}^\prime\vert$) follows from
Eq.~(\ref{eq:kktildep}) and is given by ($2
u^2=\underline{v}_\alpha\cdot \underline{v}_{\alpha}^{\prime}-1$)
\begin{equation}\label{eq:omegaif}
\omega_{\tilde{k}_{\bar{q}}}=2 \tilde{k}^{\prime 1}_{\bar{q}}\, u +2
\tilde{k}^{\prime 3}_{\bar{q}}\, u^2+
\omega_{\tilde{k}^\prime_{\bar{q}}} (2u^2+1)\, .
\end{equation}
The spin-rotation factor $\mathcal{S}_{\mathrm{IF}}$ takes on the
form
\begin{equation}
\mathcal{S}_{\mathrm{IF}}=\frac{m_{\bar{q}}+\omega_{\tilde{k}^\prime_{\bar{q}}}+\tilde{k}^{
\prime 1}_{\bar{q}}\,
u}{\sqrt{(m_{\bar{q}}+\omega_{\tilde{k}_{\bar{q}}})
(m_{\bar{q}}+\omega_{\tilde{k}^\prime_{\bar{q}}})}}\, .
\end{equation}

In the infinite-momentum frame the meson moves with large velocity
in $z$-direction and the momentum is transferred in transverse
direction. It is a special $q^+=0$ frame, in which the plus
component of the 4-momentum transfer vanishes. Such frames are very
popular for form-factor studies in
front-form~\cite{Keister:1991sb,Simula:1996pk}. Another widely used
frame to analyze the $\gamma^\ast M_\alpha\rightarrow M_\alpha$
subprocess is the Breit frame in which the energy-transfer between
the meson in the initial and the final states
vanishes~\cite{lev:1995,Klink:1998qf}. This corresponds to elastic
electron-meson backward scattering in the (overall)
center-of-momentum frame and is characterized by the minimal meson
momentum necessary for reaching a particular momentum transfer $Q$.
In this sense it is just the opposite situation to the
infinite-momentum frame, in which the meson momentum goes to
infinity. In our case the Breit frame is reached by taking the
minimum value for $\nu_\alpha$, i.e.
$\nu_\alpha^2=u^2=(\underline{v}_\alpha\cdot
\underline{v}_{\alpha}^{\prime}-1)/2$ (cf. Eq.~(\ref{eq:vscatt})).
If this is done,
\begin{eqnarray}
\tilde{J}^{\nu}_\infty(\vec{\underline{v}}_\alpha^\prime,
\vec{\underline{v}}_\alpha) &\stackrel{\nu_\alpha \rightarrow
u}{\longrightarrow}& (\underline{v}_\alpha+
\underline{v}_{\alpha}^{\prime})^\nu\,
\Big\{\tilde{f}(\underline{v}_\alpha\cdot
\underline{v}_{\alpha}^{\prime},\nu_{\alpha}=u) \nonumber\\
&& + \sqrt{\frac{\underline{v}_\alpha\cdot
\underline{v}_{\alpha}^{\prime}-1}{\underline{v}_\alpha\cdot
\underline{v}_{\alpha}^{\prime}+1}}
\,\tilde{g}(\underline{v}_\alpha\cdot
\underline{v}_{\alpha}^{\prime},\nu_{\alpha}=u)\Big\}\nonumber\\&&=:
(\underline{v}_\alpha+ \underline{v}_{\alpha}^{\prime})^\nu\,
\xi_{\mathrm{B}}(\underline{v}_\alpha\cdot
\underline{v}_{\alpha}^{\prime})
\end{eqnarray}
and it is not possible any more to separate the physical form factor
$\tilde f$ from the unphysical form factor $\tilde g$. We therefore
denote the resulting combination that occurs as coefficient of the
covariant $(\underline{v}_\alpha+
\underline{v}_{\alpha}^{\prime})^\nu$ by
$\xi_{\mathrm{B}}(\underline{v}_\alpha\cdot
\underline{v}_{\alpha}^{\prime})$, i.e. the Isgur-Wise function in
the Breit frame. The integral for
$\xi_{\mathrm{B}}(\underline{v}_\alpha\cdot
\underline{v}_{\alpha}^{\prime})$ has the same structure as the one
for $\xi_{\mathrm{IF}}(\underline{v}_\alpha\cdot
\underline{v}_{\alpha}^{\prime})$ (cf. Eq.~(\ref{eq:xiif})), namely
\begin{equation}
\xi_{\mathrm{B}}(\underline{v}_\alpha\cdot\underline{v}_{\alpha}^{\prime})=
\int\, \frac{d^3\tilde{k}_{\bar{q}}^\prime}{4\pi}\,
\sqrt{\frac{\omega_{\tilde{k}_{\bar{q}}}}
{\omega_{\tilde{k}^\prime_{\bar{q}}}}}\,\mathcal{S}_{\mathrm{B}}\,
\psi^\ast\,(\vert \vec{\tilde{k}}_{\bar{q}}^\prime\vert)\,  \psi
 \,(\vert \vec{\tilde{k}}_{\bar{q}}\vert)\,\, .
\end{equation}
Only the boosts that relate $\tilde{k}_{\bar{q}}^\prime$ and
$\tilde{k}_{\bar{q}}$ are different. In the Breit frame
$\omega_{\tilde{k}_{\bar{q}}}$ and
$\omega_{\tilde{k}^\prime_{\bar{q}}}$ are connected via
\begin{equation}\label{eq:omegabreit}
\omega_{\tilde{k}_{\bar{q}}}=2 \tilde{k}^{\prime 1}_{\bar{q}}\, u
\sqrt{u^2+1}+ \omega_{\tilde{k}^\prime_{\bar{q}}} (2u^2+1)
\end{equation}
and the spin-rotation factor $\mathcal{S}_{\mathrm{B}}$ becomes
\begin{equation}
\mathcal{S}_{\mathrm{B}}=\frac{m_{\bar{q}}+\omega_{\tilde{k}^\prime_{\bar{q}}}+\tilde{k}^{
\prime 1}_{\bar{q}}\,
\frac{u}{\sqrt{u^2+1}}}{\sqrt{(m_{\bar{q}}+\omega_{\tilde{k}_{\bar{q}}})
(m_{\bar{q}}+\omega_{\tilde{k}^\prime_{\bar{q}}})}}\, .
\end{equation}

The integrands for the Isgur-Wise function in the infinite-momentum
frame and the Breit frame are thus obviously different.
Surprisingly, the numerical integration gives the same
results for
$\xi_{\mathrm{IF}}(\underline{v}_\alpha\cdot\underline{v}_{\alpha}^{\prime})$
and
$\xi_{\mathrm{B}}(\underline{v}_\alpha\cdot\underline{v}_{\alpha}^{\prime})$.
This can be seen in Fig.~\ref{fig:vdep}, where the results for
$\xi_{\mathrm{B}}(\underline{v}_\alpha\cdot\underline{v}_{\alpha}^{\prime})$
are indicated by the black dots. These dots should be compared with the
right end of the corresponding curves. This suggests that the
integrands of
$\xi_{\mathrm{IF}}(\underline{v}_\alpha\cdot\underline{v}_{\alpha}^{\prime})$
and
$\xi_{\mathrm{B}}(\underline{v}_\alpha\cdot\underline{v}_{\alpha}^{\prime})$
are related by a change of integration variables. And indeed, the
expressions for the energies in Eqs.~(\ref{eq:omegaif}) and
(\ref{eq:omegabreit}) are connected via a simple rotation:
\begin{eqnarray}
\begin{pmatrix}
\tilde{k}_{\bar{q}}^{\prime\,1}\\
\tilde{k}_{\bar{q}}^{\prime\,3}
\end{pmatrix}_\mathrm{IF}=\frac{1}{\sqrt{u^2+1}}
\begin{pmatrix}
1 & -u\\ u & 1
\end{pmatrix}
\begin{pmatrix}
\tilde{k}_{\bar{q}}^{\prime\,1}\\
\tilde{k}_{\bar{q}}^{\prime\,3}
\end{pmatrix}_\mathrm{B}\, .
\end{eqnarray}
Applying this change of variables to the spin-rotation factor
$\mathcal{S}_{\mathrm{IF}}$ one ends up with
$\mathcal{S}_{\mathrm{B}}$ plus an additional term which is an odd
function of $\tilde{k}_{\bar{q}\mathrm{B}}^{\prime\,3}$ that
vanishes upon integration. Our result for the Isgur-Wise function is
thus independent on whether we extract it in the Breit frame or in
the infinite-momentum frame. We therefore will drop the subscripts
\lq\lq IF\rq\rq\ and \lq\lq B\rq\rq. For further purposes we will
take the somewhat simpler analytical Breit-frame expression
\begin{equation}\label{eq:IWfinal}
\xi(v\cdot v^\prime)= \int\,
\frac{d^3\tilde{k}_{\bar{q}}^\prime}{4\pi}\,
\sqrt{\frac{\omega_{\tilde{k}_{\bar{q}}}}
{\omega_{\tilde{k}^\prime_{\bar{q}}}}}\,\mathcal{S}\,
\psi^\ast\,(\vert \vec{\tilde{k}}_{\bar{q}}^\prime\vert)\,  \psi
 \,(\vert \vec{\tilde{k}}_{\bar{q}}\vert)\,\, .
\end{equation}
with
\begin{equation}\label{eq:IWomega}
\omega_{\tilde{k}_{\bar{q}}}= \tilde{k}^{\prime 1}_{\bar{q}}\,
\sqrt{(v\cdot v^\prime)^2-1}+ \omega_{\tilde{k}^\prime_{\bar{q}}}\,
(v\cdot v^\prime)
\end{equation}
and
\begin{equation}\label{eq:IWspin}
\mathcal{S}=\frac{m_{\bar{q}}+\omega_{\tilde{k}^\prime_{\bar{q}}}+\tilde{k}^{
\prime 1}_{\bar{q}}\, \sqrt{\frac{{(v\cdot v^\prime)-1}} {{(v\cdot
v^\prime)+1}}}} {\sqrt{(m_{\bar{q}}+\omega_{\tilde{k}_{\bar{q}}})
(m_{\bar{q}}+\omega_{\tilde{k}^\prime_{\bar{q}}})}}\, .
\end{equation}
as our Isgur-Wise function. Here we have just reexpressed $u$ in
terms of $v\cdot v^\prime$. As one can check, the Isgur-Wise
function introduced in this way is now only a function of $v\cdot
v^\prime$ and it is correctly normalized, i.e.
\begin{equation}
\xi(v\cdot v^\prime=1)=1\, .
\end{equation}
Its independence on the heavy-quark mass $m_Q$ is one of the
consequences of heavy-quark flavor symmetry which is supposed to
hold in the heavy-quark
limit~\cite{Isgur:1989vq,Isgur:1989ed,Neubert:1993mb}.

Heavy-quark  flavor symmetry reaches even further. The heavy flavor in the final
state can be replaced by another heavy flavor without affecting the
Isgur-Wise function. The physical processes leading to such
flavor-changing heavy-to-heavy transitions are, e.g., weak decays.
Thus our next aim will be to check whether the heavy-quark limit of
the weak $B\rightarrow D$ transition current, as given in
Eq.~(\ref{eq:Jwkpsps}), provides the same Isgur-Wise function as the
electromagnetic current, Eq.~(\ref{eq:JQ}).

\subsection{Time-like momentum transfer}
Like in the electromagnetic case we rewrite meson and heavy-quark
momenta in terms of velocities. The meson momenta that specify our
decay kinematics (cf. Eq.~(\ref{eq:decaykinem})) can be directly
expressed in terms of $\underline{v}_B \cdot
\underline{v}^\prime_{D^{(\ast)}}$:
\begin{eqnarray}\label{eq:vdecay}
\underline{k}_B&=&m_B \left( \begin{array}{c} 1\\ 0 \\0\\0
\end{array}\right)=m_B \underline{v}_B\, , \\
\underline{k}_{D^{(\ast)}}^\prime&=& m_{D^{(\ast)}}\left(
\begin{array}{c}
\underline{v}_B \cdot
\underline{v}^\prime_{D^{(\ast)}}\\
\sqrt{(\underline{v}_B \cdot \underline{v}^\prime_{D^{(\ast)}})^2-1}
\\0\\0
\end{array}\right)=m_{D^{(\ast)}}\underline{v}_{D^{(\ast)}}\,
.\nonumber
\end{eqnarray}\\
For the decay the momentum transferred between the initial and the
final meson is time-like, i.e.
\begin{eqnarray}\label{eq:qdecay}
0\leq \underline{q}^2&=&(\underline{k}_B-\underline{k}_{D^{(\ast)}})^2 \\
&&\hspace{-1.5cm}=m_B^2+m_{D^{(\ast)}}^2 -2 m_B
m_{D^{(\ast)}}\underline{v}_B \cdot
\underline{v}^\prime_{D^{(\ast)}}\leq (m_B-m_{D^{(\ast)}})^2\,
.\nonumber
\end{eqnarray}
From Eqs.~(\ref{eq:vdecay}) and (\ref{eq:qdecay}) we conclude that
\begin{equation}
1\leq \underline{v}_B \cdot \underline{v}^\prime_{D^{(\ast)}}\leq
1+\frac{(m_B-m_{D^{(\ast)}})^2}{2 m_B m_{D^{(\ast)}}}\, .
\end{equation}
Note that this $v\cdot v^\prime$-interval is also accessible in
elastic electron-meson scattering for which only $v\cdot
v^\prime\geq 1$ must hold. This makes it possible to directly
compare the structure of heavy-light mesons as measured in elastic
scattering with the structure inferred from the observation of weak
decays, although these processes involve space-like and time-like
momentum transfers, respectively.

\begin{widetext}
Since the axial-vector contribution of the quark current vanishes
for pseudoscalar to pseudoscalar transitions, the heavy-quark limit
of the $B\rightarrow D$ transition current, Eq.~(\ref{eq:Jwkpsps}),
is given by
\begin{eqnarray}\label{eq:hqljBD}
J^{\nu}_{B\rightarrow D}
(\vec{\underline{k}}_D^\prime,\vec{\underline{k}}_B)
\quad\stackrel{\mathrm{h.q.l.}}{\longrightarrow}\quad \sqrt{m_B
m_D}\, \tilde{J}^{\nu}_{B\rightarrow D}
(\vec{\underline{v}}_D^\prime,\vec{\underline{v}}_B)&=& \sqrt{m_B
m_D}\, \int\, \frac{d^3\tilde{k}_{\bar{q}}^\prime}{4\pi}\,
\sqrt{\frac{\omega_{\tilde{k}_{\bar{q}}}}
{\omega_{\tilde{k}^\prime_{\bar{q}}}}}
 \,  \bigg\{\!\sum_{\mu_b,\mu_c^\prime
=\pm \frac{1}{2}}\!\!\!
\bar{q}_{\mu_c^\prime}(\vec{\underline{v}}_D^{\,\prime})\,\gamma^\nu\,
u_{\mu_b}(\vec{\underline{v}}_B) \nonumber \\ && \times\frac{1}{2}\,
D^{1/2}_{\mu_b\mu_c^\prime}\!
\left[\!R_{\mathrm{W}}\!\left(\frac{\tilde{k}^\prime_{\bar{q}}}{m_{\bar{q}}},
B_c(\underline{v}_{D}^\prime)\right)\, \, \right] \bigg\}\,
\psi^\ast\,(\vert \vec{\tilde{k}}_{\bar{q}}^\prime\vert)\,  \psi
 \,(\vert \vec{\tilde{k}}_{\bar{q}}\vert)\,\, .
\end{eqnarray}
\end{widetext}
Here we have made use of Eq.~(\ref{eq:hqlmomenta}) and the fact that
the Wigner rotation of the $c$-quark spin becomes the identity.
Exploiting the general properties of the Wigner $D$-functions and
\begin{eqnarray}
\bar{q}_{-\mu_b}(\vec{\underline{v}}_D^{\,\prime})\,\gamma^\nu\,
u_{\mu_b}(\vec{\underline{v}}_B)&=&-\left(\bar{q}_{\mu_b}(\vec{\underline{v}}_D^{\,\prime})\,
\gamma^\nu\,
u_{-\mu_b}(\vec{\underline{v}}_B)\right)^\ast\, ,\nonumber\\
\bar{q}_{\mu_b}(\vec{\underline{v}}_D^{\,\prime})\,\gamma^\nu\,
u_{\mu_b}(\vec{\underline{v}}_B)&=&\sqrt{\frac{2}{\underline{v}_B
\cdot \underline{v}^\prime_{D}+1}}\,
(\underline{v}_B+\underline{v}^\prime_{D})^\nu\, , \nonumber\\
\end{eqnarray}
it can be shown that the heavy-quark limit of the $B\rightarrow D$
transition current finally takes on the form
\begin{equation}\label{eq:Jpspshql}
\tilde{J}^{\nu}_{B\rightarrow D}
(\vec{\underline{v}}_D^\prime,\vec{\underline{v}}_B)=
(\underline{v}_B+\underline{v}^\prime_{D})^\nu \,
\xi(\underline{v}_B\cdot\underline{v}^\prime_{D})\, ,
\end{equation}
with $\xi(\underline{v}_B\cdot\underline{v}^\prime_{D})$ being the
Isgur-Wise function defined in
Eqs.~(\ref{eq:IWfinal})-(\ref{eq:IWspin}). This proves that
heavy-quark flavor symmetry is respected by our appproach to the
electroweak structure of heavy-light mesons.

Whereas the Isgur-Wise function is just the heavy-quark limit of the
electromagnetic form factor (expressed as function of $v\cdot
v^\prime$) its relation to the decay form factors $F_0$ and $F_1$ is
a little bit more complicated. By comparing Eq.~(\ref{eq:Jpspshql})
with Eq.~(\ref{eq:Jpspsphys}) it follows that~\cite{Neubert:1991xw}
\begin{equation}\label{eq:f0xi}
R\,\left[1-\frac{q^2}{(m_B+m_D)^2}\right]^{-1}\,
F_0(q^2)\stackrel{\mathrm{h.q.l.}}{\longrightarrow}\,
\xi(\underline{v}_B\cdot\underline{v}^\prime_{D})
\end{equation}
and that
\begin{equation}\label{eq:f1xi}
R\,\, F_1(q^2)\stackrel{\mathrm{h.q.l.}}{\longrightarrow}\,
\xi(\underline{v}_B\cdot\underline{v}^\prime_{D})\, ,
\end{equation}
with
\begin{equation}\label{eq:R}
R=\frac{2\sqrt{m_B m_D}}{m_B+m_D}\, .
\end{equation}
For finite heavy quark masses the deviation of the left-hand sides
of Eqs.~(\ref{eq:f0xi}) and (\ref{eq:f1xi}) from the Isgur-Wise
function $\xi(\underline{v}_B\cdot\underline{v}^\prime_{D})$ is a
measure for the amount of heavy-quark (flavor) symmetry breaking.

The heavy-quark flavor symmetry is not the only symmetry which is
recovered in the heavy-quark limit. There is also a heavy-quark spin
symmetry which has its origin in the decoupling of the heavy-quark
spin from the spin of the light degrees of freedom.
Heavy-quark spin symmetry allows to relate matrix elements involving
vector mesons with corresponding ones for pseudoscalar mesons. A
particular example is the statement that the current matrix elements
of the pseudoscalar-to-vector $B\rightarrow D^\ast$ transition are
determined by the same Isgur-Wise function as the current matrix
elements of the pseudoscalar-to-pseudoscalar $B\rightarrow D$
transition~\cite{Isgur:1989ed,Isgur:1989vq,Neubert:1993mb}.

The heavy-quark limit of the $B\rightarrow D^\ast$ transition current, Eq.~(\ref{eq:Jwkpsv}), becomes
\begin{widetext}\begin{eqnarray}\label{eq:Jhqlpsv}
& J^{\nu}_{B\rightarrow D^\ast}
(\vec{\underline{k}}_{D^\ast}^\prime,\underline{\mu}_{D^\ast}^\prime;\vec{\underline{k}}_B)& \nonumber\\ &\downarrow{\mathrm{h.q.l.}} &\nonumber\\ & \sqrt{m_B
m_D}\, \tilde{J}^{\nu}_{B\rightarrow D^\ast}
(\vec{\underline{v}}_{D^\ast}^\prime,\underline\mu_{D^\ast}^\prime;\vec{\underline{v}}_B)
& =
\sqrt{m_B
m_D}\,  \int\, \frac{d^3\tilde{k}_{\bar{q}}^\prime}{4 \pi}\,
\sqrt{\frac{\omega_{\tilde{k}_{\bar{q}}}}
{\omega_{\tilde{k}^\prime_{\bar{q}}}}} \,
 \bigg\{\!\sum_{\mu_b,\mu_c^\prime,\tilde\mu_{\bar
 q}^\prime=\pm \frac{1}{2}}
 \!\!\!\!\!\!
\bar{q}_{\mu_c^\prime}(\underline{\vec{v}}_{D^\ast}^\prime)\,\gamma^\nu\,
(1-\gamma^5) u_{\mu_b}(\underline{\vec{v}}_B)  \\ & &\quad \times
\sqrt{2} (-1)^{\frac{1}{2}-\mu_b}
C^{1\underline\mu^\prime_{\!D^\ast}}_{\frac{1}{2}\mu_c^\prime\frac{1}{2}
\tilde{\mu}_{\bar{q}}^\prime}\, D^{1/2}_{\tilde\mu_{\bar{q}}^\prime
-\mu_b}\!\left[ \!R^{-1}
_{\mathrm{W}}\!\left(\frac{\tilde{k}^\prime_{\bar q}}{m_{\bar q}},
B^{-1}_c(\underline{v}_{D^\ast}^\prime)\right)\right] \bigg\}\,
\psi^\ast_{D^{\ast}}\,(\vert \vec{\tilde{k}}_{\bar{q}}^\prime\vert)\, \psi_B
 \,(\vert \vec{\tilde{k}}_{\bar{q}}\vert)\, .\nonumber
\end{eqnarray}
It can now be verified that $\tilde{J}^{\nu}_{B\rightarrow D^\ast}
(\vec{\underline{v}}_{D^\ast}^\prime,\underline{\mu}_{D^\ast}^\prime;\vec{\underline{v}}_B)$ has the desired covariant structure~\cite{Isgur:1989ed}
\begin{eqnarray}
\label{eq:Jpsvhqlcov}
\tilde{J}^{\nu}_{B\rightarrow D^\ast}
(\vec{\underline{v}}_{D^\ast}^\prime,\underline{\mu}_{D^\ast}^\prime;\vec{\underline{v}}_B)&=&
i\,\epsilon^{\nu\alpha\beta\gamma}\, \epsilon_\alpha(m_{D^\ast} \vec{\underline{v}}_{D^\ast}^\prime,\underline{\mu}_{D^\ast}^\prime)
\,\underline{v}^{\prime}_{D^\ast\beta}\, \underline{v}_{B\gamma} \,
\xi(\underline{v}_B\cdot\underline{v}^\prime_{D^\ast})\, ,\nonumber\\
&&-\left[\epsilon^\nu(m_{D^\ast} \vec{\underline{v}}_{D^\ast}^\prime,\underline{\mu}_{D^\ast}^\prime)
\,(\underline{v}_B\cdot\underline{v}^\prime_{D^\ast}+1)- \underline{v}^{\prime\nu}_{D^\ast}\,\epsilon(m_{D^\ast} \vec{\underline{v}}_{D^\ast}^\prime,\underline{\mu}_{D^\ast}^\prime)\cdot\underline{v}_B
\right]\, \xi(\underline{v}_B\cdot\underline{v}^\prime_{D^\ast})\,,
\end{eqnarray}
\end{widetext}
with $\xi(\underline{v}_B\cdot\underline{v}^\prime_{D^\ast})$ being again the
Isgur-Wise function defined in
Eqs.~(\ref{eq:IWfinal})-(\ref{eq:IWspin}). This proves that
also heavy-quark spin symmetry is recovered in the heavy-quark limit within our approach.\footnote{With $\kappa_{D^\ast}^2= m_{D^\ast}^2 ((\underline{v}_B\cdot\underline{v}^{\prime}_{D^\ast})^2-1)$ we see that $\epsilon_\alpha(m_{D^\ast} \vec{\underline{v}}_{D^\ast}^\prime,\underline{\mu}_{D^\ast}^\prime)$ is independent of $m_{D^\ast}$ (cf. Eq.~(\ref{eq:Dpol})).}

By comparing Eq.~(\ref{eq:Jpsvhqlcov}) with Eq.~(\ref{eq:Jphyspsv})
we finally obtain the  relations between the physical $B\rightarrow
D^\ast$ decay form factors (in the heavy-quark limit) and the
Isgur-Wise function~\cite{Neubert:1991xw}:
\begin{equation}\label{eq:a1xi}
R^\ast\,\left[1-\frac{q^2}{(m_B+m_D^\ast)^2}\right]^{-1}\,
A_1(q^2)\stackrel{\mathrm{h.q.l.}}{\longrightarrow}\,
\xi(\underline{v}_B\cdot\underline{v}^\prime_{D^\ast})\, ,
\end{equation}
\begin{equation}\label{eq:vxi}
R^\ast\,\, V(q^2)\stackrel{\mathrm{h.q.l.}}{\longrightarrow}\,
\xi(\underline{v}_B\cdot\underline{v}^\prime_{D^\ast})\, ,
\end{equation}
and
\begin{equation}\label{eq:a02xi}
R^\ast\,\, A_i(q^2)\stackrel{\mathrm{h.q.l.}}{\longrightarrow}\,
\xi(\underline{v}_B\cdot\underline{v}^\prime_{D^\ast})\, ,\quad i=0,2\, ,
\end{equation}
with
\begin{equation}\label{eq:Rs}
R^\ast=\frac{2\sqrt{m_B m_{D^\ast}}}{m_B+m_{D^\ast}}\, .
\end{equation}
If the left-hand sides of Eqs.~(\ref{eq:a1xi})-(\ref{eq:a02xi}) are
calculated with  physical heavy-quark masses, their deviation from
the Isgur-Wise function on the right-hand sides and the differences
amongst each other can be taken as a measure for the amount of
heavy-quark spin symmetry breaking.

\section{\label{sec:results}Numerical studies}
At this point we want to emphasize that the aim of this paper is not
to give quantitative  predictions for electroweak heavy-light
(transition) form factors based on a sophisticated constituent-quark
model. It is rather our intention to demonstrate that the kind of
relativistic coupled-channel approach that we are using to identify
the electroweak structure of few-body bound states is general enough
to provide also sensible results for heavy-light systems. First we
note that the electromagnetic and weak currents are solely
determined by the bound-state wave function and the
constituent-quark masses (cf. Eqs.~(\ref{eq:JQ}), (\ref{eq:Jwkpsps})
and (\ref{eq:Jwkpsv})). For our numerical studies we adopt the
simple harmonic-oscillator wave function already introduced in
Eq.~(\ref{eq:wavefunc}) and the oscillator and mass parameters
quoted there. In order to calculate the weak transition form factors
from the currents one also needs the meson masses calculated from
the harmonic-oscillator confinement potential (cf. Eqs.
(\ref{eq:F0}),(\ref{eq:F1}) and (\ref{eq:V})-(\ref{eq:A2})). We take
the physical masses, since the theoretically calculated spectrum can
always be shifted by adding an appropriate constant to the
confinement potential such that the experimentally measured
pseudoscalar and vector-meson ground-state masses (which we deal
with) are reproduced.

\begin{figure}[t!]
\includegraphics[width=0.45\textwidth]{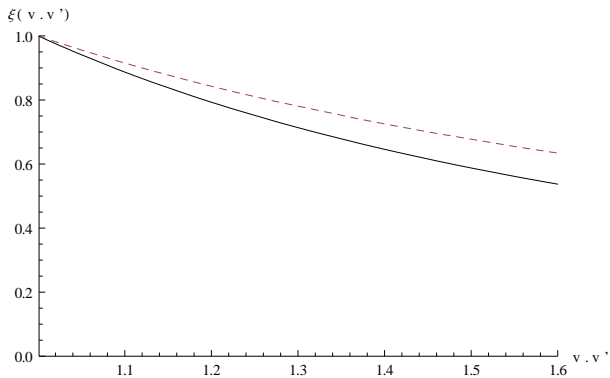}
\caption{(Color online) Isgur-Wise function (solid line) calculated
by  means of Eqs.~(\ref{eq:IWfinal})-(\ref{eq:IWspin}) with the
model parameters being the same as in Fig.~\ref{fig:sdep}. The
dashed line corresponds to spin-rotation factor
$\mathcal{S}=1$.}\label{fig:IW}\end{figure}

\begin{figure*}[ht!]
\includegraphics[width=0.45\textwidth]{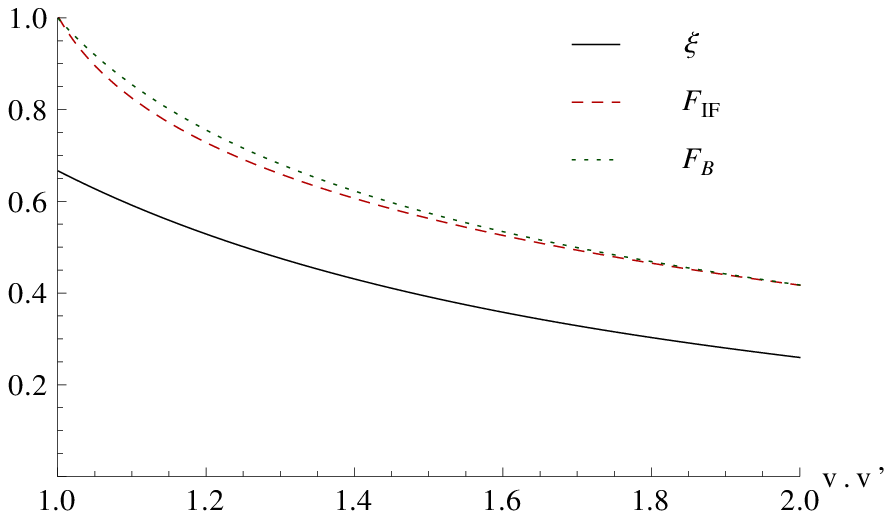}\hspace{1cm}
\includegraphics[width=0.45\textwidth]{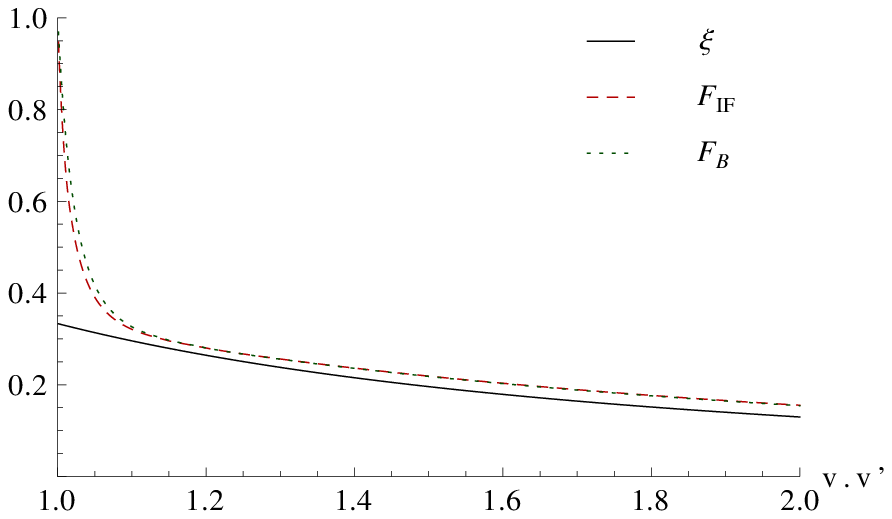}
\caption{(Color online) Electromagnetic form factors of the $D^+$
(left) and $B^-$ (right) mesons calculated in the Breit frame
(dotted line) and infinite-momentum frame (dashed line) in
comparison with the Isgur-Wise function (solid line). For direct
comparison the Isgur-Wise function is multiplied by $|Q_Q|$, i.e.
the charge of the heavy quark. Model parameters are the same as in
Fig.~\ref{fig:sdep}. } \label{fig:scattpsps}
\end{figure*}

\begin{figure*}[ht!]
\includegraphics[width=0.45\textwidth]{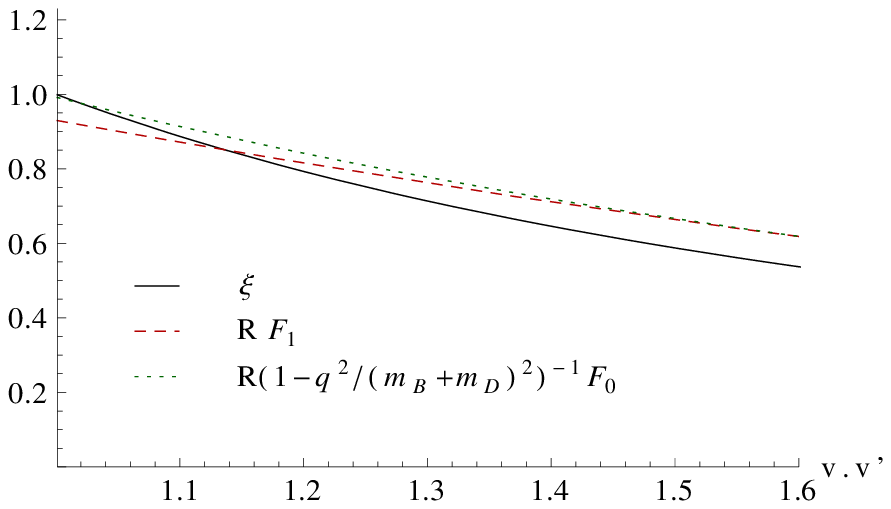}\hspace{1cm}
\includegraphics[width=0.45\textwidth]{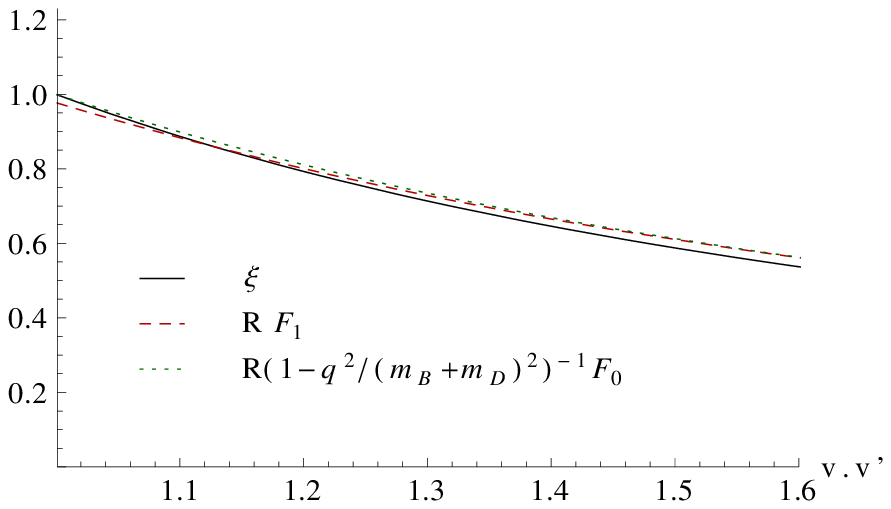}
\caption{(Color online) Weak $B^- \rightarrow D^0$
decay form factors  (multiplied with appropriate kinematical
factors, cf. Eqs. (\ref{eq:f0xi}) - (\ref{eq:R})) for physical
heavy-quark masses in comparison with the Isgur-Wise function (left
figure). Model parameters are the same as in Fig.~\ref{fig:sdep}. In
the right figure $c$ and $b$-quark masses are multiplied by a factor
$6.25$ such that $m_c=10$~GeV and meson masses are taken equal to
the corresponding quark masses.} \label{fig:decpsps}
\end{figure*}

\begin{figure*}[ht!]
\includegraphics[width=0.45\textwidth]{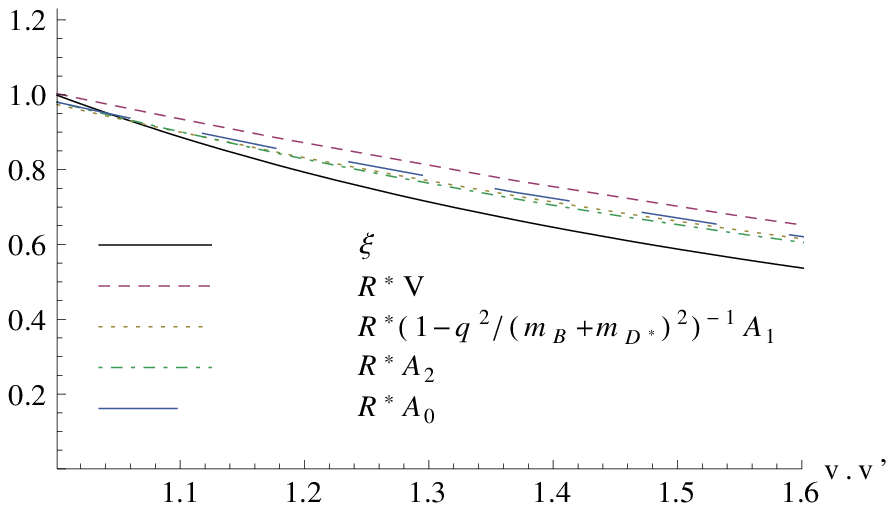}\hspace{1cm}
\includegraphics[width=0.45\textwidth]{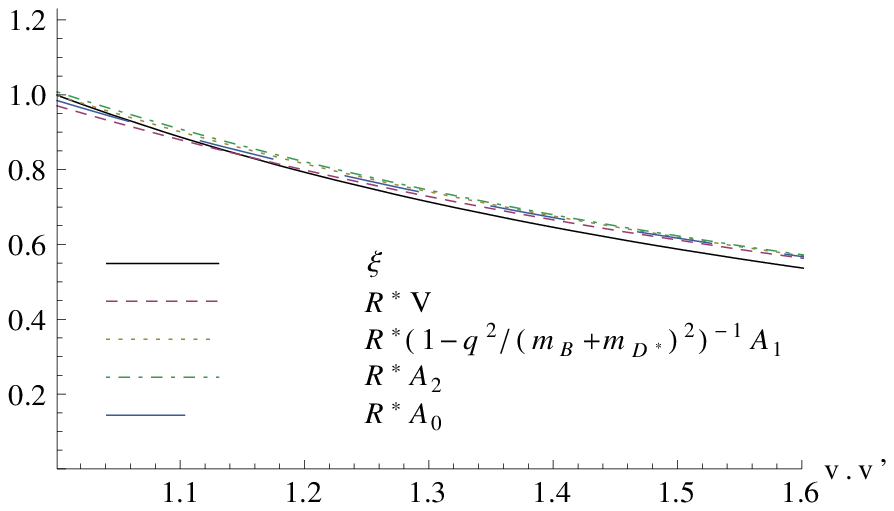}
\caption{(Color online) Weak $B^- \rightarrow D^{0\ast}$ decay form factors  (multiplied with appropriate
kinematical factors, cf. Eqs. (\ref{eq:a1xi}) - (\ref{eq:Rs})) for
physical heavy-quark masses in comparison with the Isgur-Wise
function (left figure). Model parameters are the same as in
Fig.~\ref{fig:sdep}. In the right figure $c$ and $b$-quark masses
are multiplied by a factor $6.25$ such that $m_c=10$~GeV and meson
masses are taken equal to the corresponding quark masses.}
\label{fig:decpsv}
\end{figure*}

The Isgur-Wise function, as resulting from this simple
harmonic-oscillator model,   is plotted in Figure~\ref{fig:IW}. The
effect of the quark spin onto the Isgur-Wise function can be
estimated by comparing the solid with the dashed line. The latter
corresponds to the coupling of the photon to spinless quarks and is
obtained by setting the spin-rotation factor $\mathcal{S}=1$. The
comparison shows the importance of the proper relativistic treatment
of the spin rotation when boosting the $Q$-$\bar{q}$ bound-state
wave function from the initial to the final state. Here it should be
emphasized that it does not matter within  our approach whether the
Isgur-Wise function is taken as the heavy-quark limit of the
electromagnetic $B$-meson form factor or as the heavy-quark limit of
any of the $B\rightarrow D^{(\ast)}$ decay form factors, although
these processes involve space- and time-like momentum transfers,
respectively. In the foregoing section this is proved analytically,
but it can also be verified numerically (see the right plots in
Figs.~\ref{fig:scattpsps}-\ref{fig:decpsv}). The authors of
Ref.~\cite{Cheng:1996if}, from which we have taken our model
parameters, have derived two different analytical expressions for
the Isgur-Wise function within a front-form approach by taking the
heavy-quark limit of the $B\rightarrow D$ and $B\rightarrow
D^{\ast}$  decay form factors, respectively. These two expressions
are then seen to provide the same numerical results for the Gaussian
wave function which we also use, but give different results for the
flavor dependent Wirbel-Stech-Bauer wave
function~\cite{Wirbel:1985ji}. From this they conclude that the
Wirbel-Stech-Bauer wave function violates heavy-quark symmetry. Our
numerical results, obtained with the Gaussian wave function, agree
with those of Ref.~\cite{Cheng:1996if} and we are also able to
reproduce the value for the slope of the Isgur-Wise function at the
normalization point $v\cdot v^\prime=1$, namely
$\rho^2=-\xi^\prime(1)=1.24$.

A reasonably simple analytical expression for the Isgur-Wise
function in front form can be found in Ref.~\cite{Cheng:1997au}. Its
structure bears some resemblance to
Eqs.~(\ref{eq:IWfinal})-(\ref{eq:IWspin}), but we have not attempted
to prove the equivalence. There are, however, strong hints that such
an equivalence holds. In the case of the pion we were able to show
analytically that our electromagnetic pion form factor (for
space-like momentum transfers) is equivalent with the usual
front-form expression that results from the $+$-component of a
one-body current in a $q^+=0$
frame~\cite{Biernat:2009my}.\footnote{Note that the kinematics which
we use to extract electromagnetic form factors for space-like
momentum transfers -- Eq.~(\ref{eq:momentumscatt}) with
$\kappa_\alpha\rightarrow \infty$ to get rid of cluster problems --
corresponds to a particular $q^+=0$ frame in which the $z$-component
of the meson momentum goes to infinity, i.e. the infinite-momentum
frame of the meson.} We suppose that this equivalence extends to the
case of bound states with unequal-mass constituents and generalizes
to electroweak $M\rightarrow M^\prime$ transition form factors (for
space-like momentum transfers), although we have not tried to prove
it analytically. If this is the case, the heavy-quark limit of
electroweak heavy-light meson (transition) form factors in
front-form and point-form should also lead to the same Isgur-Wise
function.

There is still one gap in this reasoning. It refers to form factors
in  the space-like momentum-transfer region, whereas the authors of
Refs.~\cite{Cheng:1996if,Cheng:1997au} derive their Isgur-Wise
function from weak $B\rightarrow D^{(\ast)}$ decay form factors,
i.e. in the time-like momentum transfer region. It cannot be taken
for granted that the heavy-quark limit of a one-body current, like
it is used in Refs.\cite{Cheng:1996if,Cheng:1997au}, gives the same
result for the Isgur-Wise function in the space- and time-like
momentum-transfer regions. Going from space- to time-like momentum
transfers means that one has to give up the $q^+=0$ condition and,
as a consequence, $Z$-graphs (i.e.  non-valence contributions) may
become important~\cite{Simula:2002vm}. This is confirmed by an
analysis of the triangle diagram for $B\rightarrow D^{(\ast)}$
decays within a simple covariant model~\cite{Bakker:2003up}. There
it is shown that analytic continuation ($q_\perp \rightarrow i
q_\perp$) of the $B\rightarrow D^{(\ast)}$ transition form factors
calculated in a $q^+=0$ frame for space-like momentum transfers to
time-like momentum transfers leads to the same results as a direct
calculation of the $B\rightarrow D^{(\ast)}$ decay form factors in
the time-like region ($q^+ \neq 0$), provided that $Z$-graph
contributions are appropriately taken into account. The importance
of $Z$-graph contributions, however, decreases with increasing mass
of the heavy quark and is generally assumed to vanish in the
heavy-quark limit, since an infinitely heavy quark-antiquark pair
cannot be produced out of the vacuum. Thus it is most likely that
the heavy-quark limit of a one-body current formulated within
front-form dynamics gives the same result for the Isgur-Wise
function in the space- and time-like momentum-transfer regions, as
it is the case in our point-form approach.

Heavy-quark symmetry is broken for finite heavy-quark masses. But
within any reasonable theoretical model for the electroweak
structure of heavy-light hadrons the heavy-quark limit of the form
factors (multiplied with appropriate kinematical factors) should go
over into one universal function, the Isgur-Wise function. It is,
however, also interesting see what has to be expected from
experimental measurements of the form factors and to estimate how
large heavy-quark-symmetry breaking effects are for physical masses
of the heavy quarks. First we discuss our model predictions for the
electromagnetic form factors of $D^+$ and $B^-$ mesons, as measured
in the space-like momentum transfer region. Fig.~\ref{fig:scattpsps}
shows these form factors as functions of $\underline{v}\cdot
\underline{v}^\prime$ in comparison with the Isgur-Wise function.
Plotted is the full form factor, as it is measured experimentally.
This includes the two contributions in which the photon goes to the
light and the heavy quark, respectively. Only the latter survives in
the heavy-quark limit. In the electromagnetic form factor these
contributions are weighted with the charges of the corresponding
quark. For direct comparison with the Isgur-Wise function one thus
also has to multiply the Isgur-Wise function with the charge of the
heavy quark. For $\underline{v}\cdot \underline{v}^\prime
\rightarrow 1$ the contribution of the light quark provides a peak
which becomes more pronounced with increasing mass of the heavy
quark. In the case of the $B^-$-meson the heavy-quark contribution
starts to dominate at $\underline{v}\cdot \underline{v}^\prime
\gtrsim 1.1$ (which corresponds to $Q^2\gtrsim 5$~GeV$^2$) and the
$\underline{v}\cdot \underline{v}^\prime$-dependence of the form
factor resembles the one of the Isgur-Wise function with the
absolute magnitude differing by about $20\%$ in the considered
$\underline{v}\cdot \underline{v}^\prime$-range. For the $D^+$-meson
the dominance of the heavy-quark contribution sets in at about the
same momentum transfer ($Q^2\gtrsim 5$~GeV$^2$), corresponding to
$\underline{v}\cdot \underline{v}^\prime \gtrsim 1.7$ (cf.
Eq.~(\ref{eq:q2scatt})). Due to the smallness of the charm-quark
mass, the absolute magnitude of the form factor at
$\underline{v}\cdot \underline{v}^\prime\approx 2$ deviates from the
Isgur-Wise function by about $60\%$.


As we have discussed already in Sec.~\ref{subsec:emff}, wrong
cluster properties inherent in the Bakamjian-Thomas construction may
lead to an unwanted dependence of the electromagnetic form factors
on Mandelstam-$s$. Note that such an $s$-dependence does not spoil
the Poincar\'e invariance of our $1$-photon-exchange amplitude, it
rather hints at a non-locality of our photon-meson vertex. If one
does not consider the full electron-meson scattering process, but
rather the $\gamma^\ast M \rightarrow M$ subprocess, the
$s$-dependence may be reinterpreted as a frame-dependence of our
description of this subprocess. The two extreme cases are minimum
$s$ to reach a particular momentum transfer $Q^2$ and $s \rightarrow
\infty$ ($Q^2$ fixed). The first corresponds to the Breit frame, the
latter to the infinite-momentum frame of the meson, respectively. In
both cases the Lorentz structure of the electromagnetic current of a
pseudoscalar meson may be expressed in terms of the physical
covariant $(\underline{p}_\alpha+\underline{p}_\alpha^\prime)^\mu$
alone and no spurious covariant or form factor is needed. The dashed
and dotted lines in Fig.~\ref{fig:scattpsps} show the
electromagnetic form factors of the $D^+$ and $B^-$ mesons for
$s\rightarrow \infty$ (infinite-momentum frame) and $s =
m_\alpha^2+m_e^2+Q^2/2+2\sqrt{m_\alpha^2+Q^2/4}\sqrt{m_e^2+Q^2/4}$
(Breit frame), respectively. The differences are already rather
small for the $D^+$ meson, become even smaller for the $B^-$ meson
and vanish in the heavy-quark limit, as we have shown analytically
in Sec.~\ref{subsec:IWspacel}.

Semileptonic decays, involving time-like momentum transfers, are
easier to handle. The decay currents that follow from our coupled
channel approach can be expanded in terms of physical covariants
alone and the form factors depend only on the 4-momentum transfer
squared (cf. Sec.~\ref{subsec:decayff}). Plotted in
Fig.~\ref{fig:decpsps} (left) are the two transition form factors
that can be measured in the weak $B^-\rightarrow D^0 e^-
\bar{\nu}_e$ decay. These form factors are multiplied with
appropriate kinematical factors such that they go over into the
Isgur-Wise function when taking the heavy-quark limit. One
prediction of heavy-quark symmetry is the approximate equality of $R
F_1$ and $R (1-q^2/(m_B+m_D)^2) F_0$. For physical masses of the
heavy quarks the differences are indeed less than $7\%$ of the
absolute values of the form factors and tend to become smaller with
increasing $\underline{v}\cdot \underline{v}^\prime$. Similar to the
case of the space-like form factor of the $B^-$ meson the deviation
from the Isgur-Wise function is still about $15\%$. In order to
demonstrate numerically that $R F_1$ and $R (1-q^2/(m_B+m_D)^2) F_0$
converge to the Isgur-Wise function in the heavy-quark limit, we
have made a calculation with $b$- and $c$-quark masses that are
$6.25$ times larger than the physical masses (such that
$m_c=10$~GeV). The result is shown in the right plot of
Fig.~\ref{fig:decpsps}. For such large masses of the heavy quark the
discrepancy between $R F_1$, $R (1-q^2/(m_B+m_D)^2) F_0$ and $\xi$
shrinks already to less than $10\%$. A quantity that is often quoted
is the slope of the Isgur-Wise function at zero recoil,
$\rho^2=-\xi^\prime(1)$. For our simple wave function model we we
have found $\rho^2=-\xi^\prime(1)=1.24$. This should be compared
with the slope of $F_D(w)=R F_1(q^2(w))$ at zero recoil,
$w=\underline{v}\cdot \underline{v}^\prime=1$, a quantity which is
directly related to the (unpolarized) semileptonic decay rate,
$d\Gamma_{B\rightarrow D e\bar{\nu}}/dw \propto (w^2-1)^{3/2}
|F_D(w)|^2$~\cite{Neubert:1993mb}. In our case we get $F_D(1)=0.93$
and $\rho^2_D:=-F_D^\prime(1)/F_D(1)=0.59$, i.e. a considerably
smaller slope than one would get in the heavy-quark limit. The up to
date experimental value for the slope, as quoted by the Heavy Flavor
Averaging Group~\cite{Asner:2010qj}, is $\rho^2_D=1.18\pm0.06$.
Combining the results for the electromagnetic form factor of the
$B^-$-meson in the space-like region and for the weak
$B^-\rightarrow D^0$ decay form factors in the time-like region one
can say that the breaking of heavy-quark flavor symmetry due to the
finite masses of the heavy quarks is at most a $15-20\%$ effect.

Similar quantitative conclusions can be drawn for the breaking of heavy quark spin symmetry from the comparison of the weak $B^-\rightarrow D^{0\ast}$ decay form factors amongst each other and with the Isgur-Wise function. Heavy-quark symmetry predicts that $R^\ast V$, $R^\ast A_0$, $R^\ast A_2$, and $R^\ast (1-q^2/(m_B+m_D)^2) A_1$ should coincide in the heavy-quark limit. The maximum difference is again about $5\%$ of the absolute value, whereas the maximum deviation from the Isgur-Wise function is about $20\%$, such that breaking of heavy-quark spin symmetry for physical quark masses in $B^-\rightarrow D^{0\ast} e^-\bar{\nu}_e$ amounts also to about $20\%$. The right plot in Fig.~\ref{fig:decpsv} shows how heavy-quark spin-symmetry is approximately restored if $b$- and $c$-quark masses are increased by about one order of magnitude.

At the end of this section we want to stress that our discussion of heavy-quark-symmetry breaking was restricted to effects that come from the finite mass of the heavy quarks. We have ignored effects that result from a (heavy) flavor dependence of the $B$- and $D^{(\ast)}$-meson wave functions, which would show up in more sophisticated constituent-quark models for heavy-light mesons. With a smaller oscillator parameter for the $D$-meson, as it is suggested in a front form analysis of heavy-meson decay constants~\cite{Hwang:2010hw}, one could, e.g., come closer to the experimental value for $\rho^2_D$.

\section{\label{sec:conclusion}Conclusions}
In this paper we have extended and generalized previous work on the electromagnetic structure of spin-0 and spin-1 two-body bound states consisting of equal-mass particles~\cite{Biernat:2009my,Biernat:2011mp,Biernat:2010tp}.
Working within the point form of relativistic quantum mechanics and using a constituent-quark model with instantaneous confining force we have derived electroweak current matrix elements and (transition) form factors for heavy-light mesons in the space- and time-like momentum-transfer regions. Starting point of this derivation is a multichannel formulation of the physical processes in which these form factors are measured, i.e electron-meson scattering and semileptonic weak decays. This formulation accounts fully for the dynamics of the exchanged gauge boson ($\gamma$ or $W$). Poincar\'e invariance is guaranteed by adopting the Bakamjian-Thomas construction with gauge-boson-fermion vertices taken from quantum field theory. Vector and axial-vector currents of the mesons can then be uniquely identified from the one-boson-exchange ($\gamma$ or $W$) amplitudes. These currents have already the right Lorentz-covariance properties and the electromagnetic current of any pseudoscalar meson is conserved. But wrong cluster properties, inherent in the Bakamjian-Thomas construction~\cite{Keister:1991sb}, give rise to spurious dependencies of the electromagnetic current on the electron momenta. For pseudoscalar mesons this unwanted dependencies are eliminated by taking the invariant mass of the electron-meson system large enough~\cite{Biernat:2009my,Biernat:2011mp,Biernat:2010tp}. The resulting electromagnetic form factor of a pseudoscalar meson is then equivalent to the one obtained in front form from the $+$-component of a one-body current in a $q^+=0$ frame. The weak pseudoscalar $\rightarrow$ pseudoscalar and pseudoscalar $\rightarrow$ vector transition currents are not plagued by such spurious contributions. They can be expressed in terms of physical covariants and form factors with the form factors depending on the (time-like) momentum transfer squared, as it should be. In front form one observes some frame dependence of the $B\rightarrow D^\ast$ decay form factors if they are extracted from the $+$-component of a simple one-body current~\cite{Cheng:1996if}. This is attributed to a missing non-valence (Z-graph) contribution, which makes the triangle diagram, from which the form factors are calculated, covariant~\cite{Cheng:1996if,Bakker:2003up}. In the case of the point form it is, of course, also not excluded that $Z$-graphs may play a role, but they are not necessary to ensure covariance of the current, since Lorentz boosts are purely kinematical and thus do not mix in higher Fock states.

Having derived comparably simple analytical expressions for the electromagnetic form factor of a pseudoscalar heavy-light meson and the $B\rightarrow D^{(\ast)}$ decay form factors we discussed the heavy-quark limit. We found that the decay form factors (multiplied with appropriate kinematical factors) go over into one universal function, the Isgur-Wise function, as demanded by heavy-quark symmetry. For the electromagnetic form factor we observed that the heavy-quark limit does not completely remove the spurious dependence on the electron momentum. One still has a spurious covariant and the $s$-dependence of the form factors goes over into a dependence on the (common) modulus of the incoming and outgoing 3-velocities of the heavy meson. This dependence on the modulus of the meson velocities vanishes by taking it large enough. In the limit of infinitely large meson velocities we found a rather simple analytical expression for the Isgur-Wise function which turned out to be (apart of a change of integration variables) the same as the expression which we got from the decay form factors. Interestingly, we have also got the same result for the Isgur-Wise function for the minimum value of the meson velocities that is necessary to reach a particular value of $\underline{v}\cdot\underline{v}^\prime$ (the argument of the Isgur-Wise function). For minimum velocities it is not possible to separate physical and spurious contributions since the respective covariants become proportional. The dependence of the electromagnetic pseudoscalar meson form factor on Mandelstam-$s$ and the dependence of the resulting Isgur-Wise function on the modulus of the meson velocities may be interpreted as a frame dependence of the $\gamma^\ast M\rightarrow M$ subprocess. The $s\rightarrow \infty$ (velocities $\rightarrow\infty$) limit corresponds to the infinite-momentum frame, whereas minimum $s$ (minimum velocities) corresponds to the Breit frame. Our finding thus means that it does not matter whether we calculate the Isgur-Wise function in the infinite-momentum frame or the Breit frame. In the heavy-quark limit the results are the same and agree with the heavy-quark limit of the decay form factors. Numerical agreement was also found with the front-form calculation of Ref.~\cite{Cheng:1996if}.

As a first application and numerical check of our approach we have calculated electromagnetic $D^+$- and $B^-$ form factors, the $B\rightarrow D^{(\ast)}$ decay form factors and the Isgur-Wise function with a simple (flavor independent) Gaussian wave function. For the electromagnetic $B^-$ form factor and for the $B\rightarrow D^{(\ast)}$ decay form factors the effect of heavy-quark symmetry breaking due to finite physical masses of the heavy quarks turned out be $15-20\%$. For the electromagnetic $D^+$ form factor it rather amounted to about $60\%$.

To conclude, we have presented a relativistic point-form formalism for the calculation of the electroweak structure of heavy-light mesons within constituent quark models with instantaneous confining forces. This formalism provides the electromagnetic form factor of pseudoscalar heavy-light systems for space-like momentum transfers and weak pseudoscalar-to-pseudoscalar as well as pseudoscalar-to-vector decay form factors for time like momentum transfers. It exhibits the correct heavy-quark-symmetry properties in the heavy-quark limit. Although we have not presented results, our approach is immediately applicable to semileptonic heavy-to-light transitions and it is general enough to deal with additional dynamical degrees of freedom, such that one could, e.g., account for non-valence Fock-state contributions in the mesons~\cite{Kleinhappel:2011is}.

\begin{acknowledgments}
We would like to thank E. Biernat for many helpful discussions. M.
G\'omez Rocha acknowledges the support of the \lq\lq Fond zur
F\"orderung der wissenschaftlichen Forschung in \"Osterreich\rq\rq
(FWF DK W1203-N16).
\end{acknowledgments}

\end{document}